\documentclass[fleqn,usenatbib]{mnras}

\usepackage{newtxtext,newtxmath}
\usepackage[T1]{fontenc}
\usepackage{ae,aecompl}

\usepackage{graphicx}	% Including figure files
\usepackage{amsmath}	% Advanced maths commands
\usepackage{amssymb}	% Extra maths symbols

\title[Statistics of CMB-lensing-derived galaxy cluster masses]{Quantifying the statistics of CMB-lensing-derived galaxy cluster mass measurements with simulations}

\author[\'I. Zubeldia \& A. Challinor]{
\'I\~{n}igo Zubeldia$^{1,2}$\thanks{E-mail: inigo.zubeldia@ast.cam.ac.uk}
and Anthony Challinor$^{1,2,3}$\thanks{E-mail: a.d.challinor@ast.cam.ac.uk}
\\
$^{1}$Institute of Astronomy, Madingley Road, Cambridge CB3 0HA, UK\\
$^{2}$Kavli Institute for Cosmology Cambridge, Madingley Road, Cambridge CB3 0HA, UK\\
$^{3}$DAMTP, Centre for Mathematical Sciences, Wilberforce Road, Cambridge CB3 0WA, UK
}

\date{Accepted XXX. Received YYY; in original form ZZZ}

\pubyear{2019}

\begin{document}
\label{firstpage}
\pagerange{\pageref{firstpage}--\pageref{lastpage}}
\maketitle

\begin{abstract}
CMB lensing is a promising, novel way to measure galaxy cluster masses that can be used, e.g., for mass calibration in galaxy cluster counts analyses. Understanding the statistics of the galaxy cluster mass observable obtained with such measurements is essential if their use in subsequent analyses is not to lead to biased results. We study the statistics of a CMB lensing galaxy cluster mass observable for a \textit{Planck}-like experiment with mock observations obtained from an $N$-body simulation. We quantify the bias and intrinsic scatter associated with this observable following two different approaches, one in which the signal due to the cluster and nearby correlated large-scale structure is isolated, and another one in which the variation due to uncorrelated large-scale structure is also taken into account. For our first approach we also quantify deviations from log-normality in the scatter, finding them to have a negligible impact on mass calibration for our \textit{Planck}-like experiment. We briefly discuss how some of our results change for experiments with higher angular resolution and lower noise levels, such as the current generation of surveys obtained with ground-based, large-aperture telescopes.
\end{abstract}


%%%%%%%%%%%%%%%%%%%%%%%%%%%%%%%%%%%%%%%%%%%%%%%%%%

%%%%%%%%%%%%%%%%% BODY OF PAPER %%%%%%%%%%%%%%%%%%

\section{Introduction}\label{sec:intro}

As the largest gravitationally-bound structures in the Universe, galaxy clusters are powerful cosmological probes \citep{Allen2011,Pratt2019}. In particular, their abundance as a function of mass and redshift, as given by the halo mass function, depends on the assumed cosmological model and on the values its parameters. Cosmological information can therefore be extracted from the observed galaxy cluster abundance in what is known as `cluster counts' analyses. In a spatially-flat $\Lambda$CDM cosmology, this abundance is particularly sensitive to the mean matter density of the Universe, which can be parametrised by $\Omega_{\mathrm{m}}$, and to the amplitude of the matter perturbations, which can be characterised by $\sigma_8$, the root mean square of the linear density fluctuations smoothed on a scale of 8\,$h^{-1}$\,Mpc. Cluster masses are, however, not directly observable, so cluster counts studies must rely on one or several cluster observables known to scale with cluster mass and use them as `mass proxies'. These observables can come from observations across different regions of the electromagnetic spectrum and include optical richness, X-ray flux, Sunyaev--Zel'dovich (SZ) flux, and lensing mass estimates, all of which are known to trace cluster masses. In recent years, studies using such observables have yielded competitive cosmological constraints~(e.g., \citealt{Mantz2010,Hasselfield2013,Planck2016xxvii,deHaan2016,Bocquet2018,Costanzi2019,Zubeldia2019,Kirby2019}). In these analyses, scaling relations are needed in order to relate accurately the cluster observables to the cluster mass, since it is the dependence of the cluster abundance on the latter that can be theoretically predicted. These scaling relations need to be calibrated, and this is typically a difficult task which often brings in significant systematic uncertainty. Indeed, currently, the determination of cluster masses constitutes the largest source of uncertainty in cluster counts studies (see \citealt{Pratt2019} for a recent review of cluster mass calibration).

%Limiting factor: absolute mass calibration. Lensing comes to the rescue. Cite examples.

Lensing observations can be very useful in this respect, since they provide almost bias-free estimates of cluster masses. In SZ counts analyses, in which an SZ-derived observable is used both to select the sample and
as a \emph{precise} (i.e., high signal-to-noise) cluster mass proxy for each cluster, the overall mass scale can be determined through \emph{exact} (i.e., unbiased, or nearly unbiased) lensing mass estimates. These lensing-derived masses generally
either have a significantly lower signal-to-noise ratio per cluster than the SZ observable, are much more costly to obtain, or a combination of both. These lensing mass estimates can be obtained from galaxy lensing observations, a well-established practice  (e.g., as in two of the calibrations used in the baseline analysis in \citealt{Planck2016xxvii}, in \citealt{deHaan2016}, and in \citealt{Bocquet2018}), or from CMB observations, a more novel approach (as in the third calibration of the baseline analysis in \citealt{Planck2016xxvii}, and in \citealt{Zubeldia2019}).

%Novel approach: CMB lensing. Explain what it is, and advantages over galaxy lensing. Use in our other paper. Need to accurately relate measured lensing masses to true mass.

First considered in \citet{Seljak2000}, CMB lensing by clusters has attracted some interest in recent years. Several mass estimators have been developed (see, e.g., \citealt{Yoo2008,Melin2015,Raghunathan2017,Horowitz2017}), and the CMB lensing signal of galaxy clusters and, in general, halos, has been detected to moderate-to-high statistical significance in various recent works \citep{Baxter2015,Madhavacheril2015,Planck2016xxvii,Baxter2017,Raghunathan2019,Zubeldia2019,Raghunathan2019b}. CMB lensing cluster mass estimation has several intrinsic virtues, particularly in the context of SZ surveys. First, mass estimates can be obtained from the same data set as the SZ observable, and are relatively cheap to obtain, in contrast to most current galaxy-lensing-derived masses. This enables one to obtain CMB lensing mass estimates for the totality of the clusters in the SZ sample in a relatively straightforward way provided that redshift measurements of the clusters are available \citep{Planck2016xxvii,Zubeldia2019}, as opposed to most current galaxy lensing mass determinations, in which typically the masses of only a small subset of clusters in the SZ sample are obtained. Moreover, the signal-to-noise of CMB lensing mass estimates does not decrease strongly with redshift (see, e.g., \citealt{Melin2015}), whereas high-redshift galaxy lensing mass determinations can suffer from the lack of a sufficient number of background galaxies. In addition, the CMB lensing signal is not affected by the uncertainties in the photometric redshifts of the background galaxies, which are a limiting factor in galaxy lensing analyses. However, for an experiment like \textit{Planck}, CMB lensing mass estimates are much noisier than state-of-the-art galaxy lensing mass estimates (e.g., \citealt{Linden2014,Hoekstra2015}), with typical signal-to-noise ratios of a fraction of unity for large clusters \citep{Zubeldia2019}. Furthermore, being a more novel technique, cluster CMB lensing mass estimation methods have not been tested to the extent that galaxy lensing methods have (see, e.g., \citealt{Raghunathan2017} for a quantification of several possible systematics that can affect CMB lensing cluster mass measurements).

%What needs to be understood is P(mass proxy|M_500). If functional shape is assumed, this translates to determine parameters. Option: self-calibrating at surveys. Problems: highly degenerate with other parameters (e.g. SZ bias), and loss of constraining power. Alternative: simulations. Advantages: know true mass, calibration is accurate; can also check the functional shape. Cons: validity of sims, systematics very difficult to quantify.

A fundamental element of a cluster cosmological analysis that needs to be determined accurately in order for the analysis to deliver unbiased cosmological information is the relation between the mass observable (or observables) and the true cluster mass. This involves determining both the scaling relation that relates some mean value of the observable(s) with the cluster true mass, accounting for any possible biases, and the statistical scatter that exists around the mean value of the observable(s), which typically has a complex origin and is difficult to predict from first principles \citep{Allen2011}. More rigorously, what needs to be specified is the conditional probability density function followed by the mass observable(s), obs, conditioned on the value of the true cluster mass, $M_{500}$, $P(\mathrm{obs}|M_{500})$. We note that here and throughout we choose as a cluster's `true mass' the mass within a sphere within which the mean density is 500 times the critical density at the cluster's redshift, and we denote it with $M_{500}$. We also note that this conditional probability density function can also be conditioned on other variables, such as redshift $z$; we omit these possible additional conditioning variables here for concision, but $P(\mathrm{obs}|M_{500})$ should be thought of as potentially having more conditioning variables implicit. If a functional form is assumed for $P(\mathrm{obs}|M_{500})$, its parameters can be self-calibrated with real data. However, doing this can have a negative impact on the statistical power of the cosmological analysis, especially if the calibration is carried out at the cosmological analysis level. An alternative approach is to use cosmological simulations in order to determine $P(\mathrm{obs}|M_{500})$, or at least to inform the likely range of parameter values of some assumed functional form. This also makes it possible to assess whether the chosen functional form is a good description of $P(\mathrm{obs}|M_{500})$. However, this approach is obviously limited by the accuracy of the simulations themselves, something that is difficult to quantify.

The distribution $P(\mathrm{obs}|M_{500})$ has been widely studied in the literature from both simulations and observations for a number of cluster observables (e.g., \citealt{Pratt2009,Becker2011,Angulo2012,Bahe2012,Rasia2012,Shirasaki2016,Geach2017}). Nevertheless, it has never been studied for a CMB lensing mass observable. This is the aim of this paper: to quantify $P(\mathrm{obs}|M_{500})$, where $\mathrm{obs}$ is the CMB lensing mass observable that is used in \citet{Zubeldia2019}, the CMB lensing signal-to-noise $p_{\mathrm{obs}}$. We follow the simulation approach, producing mock observations of $p_{\mathrm{obs}}$ for all the clusters with $M_{500} > 2 \times 10^{14} M_{\odot}$ in two snapshots of a large $N$-body cosmological simulation, BigMDPL \citep{Klypin2016}. We think of $p_{\mathrm{obs}}$ as consisting of three different contributions,
\begin{equation}
p_{\mathrm{obs}} = p_{\textrm{c}} + \Delta p_{\textrm{u}} + \Delta p_{\textrm{n}}\, .
\label{eq:intro1}
\end{equation}
Here, $p_{\textrm{c}}$ is the contribution to the lensing mass observable from the cluster itself, which includes the variation due to cluster triaxiality, and from large scale structure (LSS) correlated with the cluster. The other two terms, $\Delta p_{\textrm{u}}$ and $\Delta p_{\textrm{n}}$, are the contributions originating from LSS uncorrelated with the cluster and from lensing reconstruction noise, respectively. We follow two different approaches in order to analyse our mock observations. In the first approach, which we call our \textit{deconvolution} approach, we treat both $\Delta p_{\textrm{u}}$ and $\Delta p_{\textrm{n}}$ as noise and then use our mock observations in order to characterise $p_{\textrm{c}}$. In the second approach, which we call our \textit{extrapolation} approach, we only treat $\Delta p_{\textrm{n}}$ as noise, and then characterise $p \equiv p_{\textrm{c}} + \Delta p_{\textrm{u}}$ with our mock observations, extrapolating our results to the full line-of-sight from $z=0$ back to CMB last scattering.
 
The main motivation for this work is to justify the choice of priors imposed on the parameters $1-b_{\textrm{CMBlens}}$ and $\sigma_{\textrm{CMBlens}}$, which quantify, respectively, the bias and intrinsic scatter in the CMB lensing mass measurements, in the likelihood analysis of \citet{Zubeldia2019}. In particular, this is done with our deconvolution approach, which is the one followed in the hierarchical model of \citet{Zubeldia2019}. However, we also hope that our methods to study $P(\mathrm{obs}|M_{500})$ and that our main insights may be useful for future cluster counts analyses that may use CMB lensing masses as a mass proxy, which otherwise will have to determine their appropriate conditional probability distribution $P(\mathrm{obs}|M_{500})$. Throughout this paper we consider an idealised experiment with specifications similar to those of \textit{Planck}; however, we also briefly consider how some of the results change if a different experimental set-up is considered. 

%Sections of paper

This paper is organised as follows. First, in Section \ref{sec:basics} we give a brief introduction to CMB lensing. In Section \ref{sec:sims} we describe the cosmological simulations we use in order to obtain our mock observations, and in Section \ref{sec:obs} we explain how we make such mock observations. Next, in Section \ref{sec:results} we present our results, characterising $P(\textrm{obs}|M_{500})$ for a \textit{Planck}-like experiment. In this section, we first explain our model of the mock observations in Section \ref{subsec:model}, and then in Sections \ref{subsec:a} and \ref{subsec:b} we present our results for our deconvolution and extrapolation approaches, respectively. In Section \ref{sec:different} we consider how the extrapolation approach results change if different experiment specifications are considered, and we finally conclude in Section \ref{sec:conclusions}.

\section{Basics of CMB lensing}\label{sec:basics}

Massive bodies deflect light due to the effect of their gravity, a phenomenon known as gravitational lensing. CMB photons coming from the last-scattering surface are therefore deflected, the observed net effect being a remapping of the CMB fluctuations on the sky by some deflection field $\bmath{\alpha}(\hat{\bmath{n}})$ (see \citealt{Lewis2006} for a general review of CMB lensing).  

Let $X$ be an `unlensed' CMB field, i.e., a CMB field as it would have been observed if there was no lensing, where $X$ can be $T$ (the CMB temperature), $Q$, or $U$ (the two linear polarization Stokes' parameters). Lensing remaps the CMB fields so that the lensed field $\tilde{X}(\hat{\bmath{n}})$ along the line-of-sight direction $\hat{\bmath{n}}$ is the unlensed field at $\hat{\bmath{n}} + \bmath{\alpha}(\hat{\bmath{n}})$, i.e., $\tilde{X}(\hat{\bmath{n}}) = X(\hat{\bmath{n}}+\bmath{\alpha})$. 

At leading order, the deflection field can be written as $\bmath{\alpha} = \nabla_{\hat{\bmath{n}}} \psi$, where $\nabla_{\hat{\bmath{n}}}$ denotes the angular derivative (covariant derivative on the unit sphere, or, in the flat-sky approximation, partial derivative with respect to the two local angular variables) and where $\psi$ is known as the lensing potential. For a flat universe and using the Born approximation, the lensing potential can be written as \citep{Lewis2006}
\begin{equation}\label{potentiall}
\psi (\hat{\bmath{n}}) = -\frac{2}{c^2} \int_0^{\chi_{\star}} d\chi \frac{\chi_{\star}-\chi}{\chi_{\star}\chi} \Psi (\chi \hat{\bmath{n}},\eta_0-\chi)\, ,
\end{equation}
where $\chi_{\star}$ is the comoving distance to last scattering ($\chi_{\star} \approx 14$\,Gpc), $\eta_0$ is the current conformal time, and $\Psi$ is the Newtonian gravitational potential (or, in a general relativistic framework, the Weyl potential). The lensing potential is therefore a weighted integral of the gravitational potential along the undeflected line of sight.

It is often useful to work with the lensing convergence, $\kappa$, which is given by the two-dimensional Laplacian of the lensing potential, 
\begin{equation}\label{eq:cmbconv}
\kappa (\hat{\bmath{n}}) = - \frac{1}{2} \nabla_{\hat{\bmath{n}}}^2 \psi(\hat{\bmath{n}}) = \frac{3}{2} \left( \frac{H_0}{c} \right)^2 \Omega_{\mathrm{m}} \int_{0}^{\chi_{\star}} \frac{\chi \left( \chi_{\star} - \chi \right)}{\chi_{\star}} \frac{\delta}{a} d\chi \, ,
\end{equation}
where $\delta$ is the matter density contrast, $a$ is the scale factor normalised to unity today and $H_0$ is the Hubble constant.

It can be seen that the convergence is a weighted integral of the matter overdensity along the undeflected line of sight.This integrated matter distribution can be reconstructed from the CMB through the non-Gaussian features that are imprinted by lensing. Several methods to reconstruct the lensing convergence (or, equivalently, the lensing potential) exist, the most computationally simple being based on quadratic estimators \citep{Hu2001,Hu2002}.

Lensing by galaxy clusters produces variations typically of order $10\,\mu$K in the measured CMB temperature (e.g.,~\citealt{Lewis2006}), which are large enough to be probed by experiments like \textit{Planck} in a statistical way. If a cluster density profile is assumed and the cluster redshift is known, the cluster mass can be estimated by, e.g., fitting an expected cluster convergence profile to a non-parametrically reconstructed convergence, or by fitting the cluster model directly to the CMB maps.

\section{Simulation and convergence maps}\label{sec:sims}

%Basic description: cosmology, some technical details

In our study we use two snapshots of the BigMDPL simulation, a large, state-of-the-art simulation part of the publicly-available MultiDark simulation suite \citep{Klypin2016}. It consists of $3840^3$ particles of mass $2.4\times 10^{10} h^{-1} M_{\odot}$ in a simulation box of 2.5$h^{-1}$\,Gpc with periodic boundary conditions and evolved in a flat $\Lambda$CDM cosmology with $H_0 = 67.8$\,km\,s$^{-1}$\,Mpc$^{-1}$, $\Omega_{\textrm{m}} = 0.307$, baryon density parameter $\Omega_{\textrm{b}} = 0.048$, $\sigma_8 = 0.829$, and scalar spectral index $n_s = 0.96$. We consider the snapshots at $z = 0.23$ and $z=0.52$.

%Find halos, measure M_500. 

We use the halo positions as determined with the BDM (`Bound Density Maximum') technique \citep{Klypin1997,Riebe2013} that are publicly available in the CosmoSim database\footnote{\texttt{www.cosmosim.org}}. We then measure the spherical overdensity mass of each halo, $M_{500}$, using as density contrast $\Delta=500$ defined with respect to the critical density. We note that we do not remove unbound particles in this process. We refer to $M_{500}$ as the halo (or cluster) `true mass'. We only consider clusters with $M_{500} > 2\times10^{14} M_{\odot}$ and that are not subhalos of another halo, which yields a cluster catalogue with a total of $60\,391$ clusters in the $z = 0.23$ snapshot and of $29\,561$ clusters in the $z = 0.52$ snapshot. This cluster range corresponds to the most massive clusters of the Universe and spans most of the clusters in the \textit{Planck} MMF3 cosmology sample, the sample used in the cosmological analysis in \citet{Planck2016xxvii} and in \citet{Zubeldia2019}; see \citet{Planck2016xxvii} for how this sample is constructed.

%Convergence for 7 boxes centred on halos.

In order to obtain our CMB lensing measurements, the key quantity that needs to be produced is the lensing convergence around the location of each cluster in our catalogue and along a given direction of observation. We generate convergence maps using the Born approximation; under this approximation, for a spatially-flat cosmology the lensing convergence can be obtained with Eq. (\ref{eq:cmbconv}).

We produce square convergence maps centred at the location of each cluster in our catalogue for both snapshots. Each map has an angular size of 128\,arcmin along both directions perpendicular to the line of sight, and we always take the  $z$-direction, as defined within the simulation, as our line-of-sight direction. Obviously, with our simulation snapshots we cannot compute the total lensing convergence from $\chi=0$ all the way back to last scattering, $\chi=\chi_{\star}$, as in Eq. (\ref{eq:cmbconv}). We instead restrict our integration limits to a relatively small interval of comoving distance. Specifically, we take each cluster to be at the comoving distance given by the redshift of the snapshot to which it belongs, $\chi_{c}$, and we then consider seven sets of integration limits in Eq. (\ref{eq:cmbconv}), all of them centred at $\chi_{c}$ but with an increasing comoving length $\chi_l$, $\left[ \chi_{c} - \chi_l/2, \chi_{c} + \chi_l/2 \right]$. We take $\chi_l$, which hereafter we will refer to as `integration length', to be logarithmic spaced between $5$\,Mpc and $400$\,Mpc, thus taking values of 5, 10.4, 21.5, 44.7, 92.8, 192.7, and 400\,Mpc (comoving). We do this in order to investigate how the statistics of our CMB lensing mass observable depend on the relative amount of large scale structure (LSS) correlated and uncorrelated  with the cluster that is present in the convergence maps. We note that the way in which we compute our convergence maps from the simulation snapshots neglects the temporal evolution of $\delta$ along the line of sight, since in a snapshot all the particle positions are given at the same time. The error we introduce by doing this should be negligible for the relatively small integration lengths (on cosmic scales) that we are considering. We do, however, take into account the change of the geometrical lensing kernel in Eq.~(\ref{eq:cmbconv}) along the line of sight.

To summarise, for each cluster in our catalogue at any of the two considered redshifts we compute seven convergence maps, each of which corresponds to a different integration length. We refer to these maps as `cluster convergence maps' at redshift $z$ with integration length $\chi_l$.

%Random maps

Furthermore, for each of the two snapshots we compute $10^5$ square convergence maps centred at random locations in the simulation box, with the same angular size as the cluster convergence maps (128\,arcmin across each direction), and with the two largest integration lengths used in the computation of the cluster convergence maps. We refer to these maps as `random convergence maps' at redshift $z$ with integration length $\chi_l$, and describe their interpretation and use in Section \ref{sec:results}.

\section{Mock observations}\label{sec:obs}

In this section we describe our CMB lensing cluster mass observable and how we obtain mock measurements from our cluster convergence maps in order to produce a mock data set, which we then analyse in Section \ref{sec:results}.

We think of each cluster convergence map with a given integration length as the `true' convergence map around the location of that cluster due to the mass distribution contained within the considered integration length, and we denote it with $\kappa_{\mathrm{c}}$. As mentioned in Section \ref{sec:basics}, a fixed lensing convergence remaps the CMB anisotropies in a way that makes it possible for it to be estimated (or `reconstructed') from CMB observations alone. Throughout this work, except in Section \ref{sec:different}, we consider an idealised \textit{Planck}-like CMB experiment with Gaussian beam with full-width at half maximum (FWHM) of 5\,arcmin and with Gaussian instrumental noise with temperature noise levels of 45\,$\mu$K\,arcmin. We do not study the impact of uncleaned or residual foregrounds on our results. As already noted above, in Section \ref{sec:different} we investigate how some of our results change with different experimental specifications.

We consider the simplest of the CMB lensing reconstruction techniques, the quadratic estimators \citep{Hu2001,Hu2002}, thus called because they are quadratic in the CMB fields. Specifically, we only use the $TT$ quadratic estimator, which produces a reconstructed convergence map out of two copies of a CMB temperature map. We make this choice because this estimator is close to the optimal lensing estimator for a \textit{Planck}-like experiment. Using the flat-sky approximation, which is accurate given the small angular size of our convergence maps, the reconstructed convergence map that corresponds to one of our cluster convergence maps, $\hat{\kappa}$, can be written in Fourier space as
\begin{equation}\label{eq:recon}
\hat{\kappa} (\mathbf{L}) = \kappa_{\mathrm{c}} (\mathbf{L}) + n (\mathbf{L})\, .
\end{equation}
Here, $\kappa_{\mathrm{c}}(\mathbf{L})$ is the Fourier transform of the `true' convergence (the cluster convergence map), and $n (\mathbf{L})$ is the Fourier transform of the reconstruction noise, which in our simulated experiment has contributions from the primary CMB fluctuations and from instrumental noise. This noise is approximately Gaussian \citep{Lewis2006}. Following \citet{Hu2007}, in our quadratic estimator implementation we impose a Fourier-space top-hat low-pass filter on the gradient leg of the estimator, zeroing the gradient map for modes with $L > 2000$. This ensures that the quadratic estimator remains unbiased around regions of large convergence, e.g., in the central regions of galaxy clusters. In addition, as is customary, we use the lensed CMB power spectrum in the weighting of the gradient leg of the estimator and in the normalisation, rather than the unlensed CMB power spectrum, as this gives (approximately) the correct response of the estimator $\hat{\kappa}(\bmath{L})$ to lenses at wavevector $\bmath{L}$ averaged over all other lensing modes~\citep{Lewis2011,Hanson2011}. This is the same form of the quadratic estimator that we used in our analysis of real \textit{Planck} data in \citet{Zubeldia2019}. It provides, to a good approximation, unbiased reconstructed convergence maps, i.e., $\left\langle \hat{\kappa} (\mathbf{L}) \right\rangle =  \kappa_{\mathrm{c}} (\mathbf{L})$, where angular brackets denote ensemble averaging over reconstruction noise, that is, over CMB and instrumental noise fluctuations.

In our study we do not produce reconstructed maps from our cluster convergence maps, $\kappa_{\mathrm{c}}$, but instead we directly use the cluster convergence maps themselves. The reasons for doing this are the following. First, as just noted, to a good approximation our lensing reconstruction method produces unbiased convergence estimates for the experimental specifications considered. In addition, the scatter of the final lensing mass observable arising from reconstruction noise is already well understood. Indeed, since to a good approximation the reconstruction noise is Gaussian, it is fully described by its variance, which can be written as
\begin{equation}
\left\langle  \hat{\kappa} \left(\bmath{L}\right)  \hat{\kappa}^\ast \left(\bmath{L}'\right) \right\rangle = \delta^{(2)}  \left(\bmath{L} - \bmath{L}' \right) N_{\kappa}(\bmath{L})\, ,
\end{equation}
where $N_{\kappa}(\bmath{L})$ is well approximated in our case by $N^{(0)} (L)$, the normalisation of the quadratic estimator (see e.g., \citealt{Hu2001} for an analytic expression). As described below, our mass observable is linear in the convergence; thus, its scatter due to reconstruction noise is also Gaussian to a good approximation.

The next step of our measurement pipeline is to match-filter our cluster convergence maps with a cluster convergence model in order to obtain a mass observable measurement for each cluster. We use exactly the same matched-filter implementation that was used in \citet{Zubeldia2019}, and which follows \citet{Melin2015}. As a cluster model we adopt a truncated NFW profile \citep{Navarro1997}
\begin{equation}
   \rho(r) = 
\begin{cases}\label{cluster}
    \frac{\rho_0}{(r/r_s)(1 + r/r_s)^2} & \text{if } r\leq R_{\textnormal{trunc}} \, , \\
    0              & \text{if } r > R_{\textnormal{trunc}} \, , \\
\end{cases}
\end{equation}
where $r_s$ is a characteristic scale radius, $R_{\textnormal{trunc}}$ is the truncation radius, and $\rho_0$ is a characteristic density, given by
\begin{equation}
\rho_0 = \rho_c(z) \frac{500}{3} \frac{c_{500}^3}{\ln(1 + c_{500}) - c_{500}/(1+c_{500})} \, ,
\end{equation}
where $\rho_c(z)$ is the critical density at the cluster's redshift, and $c_{500}$ is the concentration parameter, which is defined as $c_{500} = R_{500} / r_s$. As in \citet{Zubeldia2019}, we choose $R_{\textnormal{trunc}} = 5 R_{500}$ and we fix  $c_{500} = 3$. Thus, in our model, a cluster is completely specified by two parameters, e.g., $M_{500}$ and redshift, $z$. We use this parameterisation throughout this paper, and refer to $M_{500}$ and $z$ as the `true cluster parameters'. 

As detailed in \citet{Zubeldia2019}, we can then write the convergence due to our cluster model at an angular separation from the cluster centre $\bmath{\theta}$ as
\begin{equation}\label{template}
\kappa_{\text{m}}(\bmath{\theta}) = \kappa_0 \kappa_{\text{t}}(\bmath{\theta};\theta_s) \, .
\end{equation}
Here, $\kappa_0$ is a normalisation such that $\kappa_0 \theta_s^2$ is proportional to $M_{500}$ and $\kappa_{\text{t}}(\bmath{\theta};\theta_s)$ is a circularly-symmetric template function that depends only on $\theta/\theta_s$, where $\theta_s$ is the angular size of the scale radius $r_s$, $\theta_s = r_s/d_A(z)$, with $d_A(z)$ the angular diameter distance to the cluster. The template is normalised to unity at $\theta=\theta_s$. Given an estimate of the convergence around a cluster, $\hat{\kappa}(\bmath{L})$, and a fiducial value of the cluster angular size, $\theta_s^{\mathrm{fid}}$, an estimator for $\kappa_0$ can be written as
\begin{equation}\label{filtval}
\hat{\kappa}_0 = \left[\int \frac{d^2 \bmath{L}}{2 \upi} \frac{|\kappa_{\textnormal{t}} (\bmath{L})|^2}{N_{\kappa}(\bmath{L})} \right]^{-1} \int \frac{d^2 \bmath{L}}{2 \upi} \frac{\hat{\kappa}(\bmath{L})  \kappa_{\textnormal{t}}^\ast (\bmath{L}) }{N_{\kappa}(\bmath{L})} \, ,
\end{equation}
where $N_{\kappa} (\bmath{L})$ is the variance of the reconstruction noise of $\hat{\kappa}(\bmath{L})$, and where the dependence of $\kappa_{\textnormal{t}}$ on $\theta_s^{\mathrm{fid}}$ is left implicit. This inverse-variance weighting ensures that the estimator is minimum-variance, which can be written as
\begin{equation}\label{filtvar}
\sigma_{\kappa_0}^2 \equiv \left\langle (\hat{\kappa}_0 - \kappa_0)^2 \right\rangle = \frac{1}{2\upi} \left[\int \frac{d^2 \bmath{L}}{2 \upi} \frac{|\kappa_{\textnormal{t}} (\bmath{L})|^2}{N_{\kappa}(\bmath{L})} \right]^{-1} \, .
\end{equation}
We use each cluster's true mass, $M_{500}$, as the model fiducial mass, from which we derive its fiducial angular scale $\theta_s^{\mathrm{fid}}$:
\begin{equation}\label{massdefinition}
M^{\mathrm{fid}} = 500 \frac{4\pi}{3} \left[d_A(z) c_{500} \theta_s^{\mathrm{fid}}\right]^2 \rho_{c}(z) \, .
\end{equation}

As discussed in \citet{Zubeldia2019}, from this estimator $\hat{\kappa}_0$ a mass estimator $\hat{M}_{500}$ can be obtained in a straightforward way, but the signal-to-noise on $\hat{\kappa}_0$ (or, equivalently, on $\hat{M}_{500}$) turns out to have better properties as an observable. This is defined as
\begin{equation}
p_{\textrm{obs}} \equiv \hat{\kappa}_0 / \sigma_{\kappa_0} \, .
\end{equation}{}
As can be seen from Eq. (\ref{filtval}), for a given cluster $p_{\textrm{obs}}$ is an unbiased estimator of the cluster CMB lensing signal-to-noise only if the convergence model, $\kappa_{\text{m}}$, is equal to the true cluster convergence, $\kappa_{\mathrm{c}}$. As shown in \citet{Zubeldia2019}, $p_{\textrm{obs}}$ is more immune to mismatch between $\kappa_{\text{m}}$ and $\kappa_{\mathrm{c}}$. In particular, it is much less dependent on the choice of $\theta_s^{\mathrm{fid}}$. We therefore use $p_{\textrm{obs}}$ as our cluster observable; $p_{\textrm{obs}}$ is also the observable used in \citet{Zubeldia2019}.

As mentioned above, the statistics of the lensing reconstruction noise are well understood, and due to this we do not produce reconstructed convergence maps of each cluster convergence map, but instead we use each cluster's true convergence, $\kappa_{\mathrm{c}}$ in place of $\hat{\kappa}$ in the matched-filtering process. That is, what we compute is the mean CMB lensing signal-to-noise, $p \equiv \left\langle p_{\textrm{obs}} \right\rangle$, where angular brackets denote averaging over reconstruction noise, $n (\mathbf{L})$, for each cluster in our catalogue and for all our seven integration lengths:
\begin{equation}\label{pdef}
p = \sqrt{2\pi} \left[\int \frac{d^2 \bmath{L}}{2 \upi} \frac{|\kappa_{\textnormal{t}} (\bmath{L})|^2}{N_{\kappa}(\bmath{L})} \right]^{-1/2} \int \frac{d^2 \bmath{L}}{2 \upi} \frac{\kappa_{\mathrm{c}}(\bmath{L})  \kappa_{\textnormal{t}}^\ast (\bmath{L}) }{N_{\kappa}(\bmath{L})} \, .
\end{equation}
%
%It is the statistics of $p$, in particular, $P(p|M_{500},z)$, that will concern us in the rest of this paper. We refer to this layer between the observed value of the observable and the true cluster parameters, $M_{500}$ and $z$, as \emph{intrinsic} scatter. $p_{\textrm{obs}}$ can then simply be thought of as a noisy realisation of $p$; we refer to this layer as \emph{observational} scatter. It has simple statistics: as mentioned above, since for our experiment specifications lensing reconstruction is very close to Gaussian, and since our observable is linear in the convergence, the observational scatter, $P(p_{\textrm{obs}}|p)$, can be well approximated by a Gaussian with, by construction, unity standard deviation.

In part of our analysis -- specifically, in Section \ref{subsec:a} -- we also treat as noise the contribution to the cluster convergence maps $\kappa_{\mathrm{c}}$ coming from large scale structure (LSS) uncorrelated with the cluster. In this case, the variance of the reconstructed convergence, after averaging over reconstruction noise and this uncorrelated LSS, is therefore $N^{(0)} (L) + C_L^{\kappa \kappa}$, where $C_L^{\kappa \kappa}$ is the convergence power spectrum. We also make measurements of $p$ for all our cluster convergence fields using this variance in the matched filter.

In summary, we obtain two sets of mock measurements of our cluster CMB lensing mass observable, $p$, for all our clusters, which are at two different redshifts, and for the seven integration lengths $\chi_l$ considered. Each set has a different understanding of what is thought of as noise in the matched-filtering process. In the first set, noise is understood as being solely due to reconstruction noise; this data set, which we refer to as cluster data set $B$, is analysed in Section \ref{subsec:b}. In the second case, the variation due to uncorrelated LSS is also included in the noise budget; this data set, which we refer to as cluster data set $A$, is studied in Section \ref{subsec:a}.

We also apply our measurement pipeline to our random convergence maps. The idea is to obtain the response of our measurement method to an observation that is not centred on a cluster, but where only the convergence due to random LSS is present. The use of these observations will become clear in Section \ref{sec:results}. For each random convergence map, we obtain a set of measurements following our measurement pipeline, using as matched-filter fiducial mass $M^{\textrm{fid}}$ the centres of 90 equally-spaced subdivisions of the interval $\text{(2--20)}\times 10^{14} M_{\odot}$, and as redshift the corresponding snapshot redshift. Our $10^5$ random convergence fields at each snapshot redshift yield $10^5$ mock measurements for each mass bin, redshift, and integration length. They constitute two sets of measurements of $p$ as a function of filter fiducial mass and integration length, which we refer to as random data set $A$ (in which the LSS power spectrum is included in the matched filter inverse-variance weighting) and $B$ (in which only the reconstruction noise power spectrum is included in the matched filter inverse-variance weighting).

\section{Statistics of the CMB lensing cluster mass observable}\label{sec:results}

In this section we analyse our mock observations, which we obtain as detailed in Section \ref{sec:obs}, in order to characterise the statistics of our CMB lensing cluster mass observable. First, in Section \ref{subsec:model}, we describe how we understand our observations in terms of random variables. We then study our mock observations following the deconvolution and extrapolation
approaches in Sections~\ref{subsec:a} and~\ref{subsec:b}, respectively.

\subsection{Model of observations}\label{subsec:model}

Let us consider our \emph{noisy} CMB lensing cluster mass observable, $p_{\mathrm{obs}}$, with either of our two choices of matched-filter inverse-variance weighting, and for a given integration length $\chi_l$. Let us also consider a point in true cluster parameter space, $M_{500}$--$z$. We can think of $p_{\mathrm{obs}}$ as a random variable, with variability arising from different sources. First, there is reconstruction noise, which arises from the CMB and instrumental noise fluctuations. In addition, there is a contribution coming from variation in the true lensing convergence. Indeed, clusters with a given true mass $M_{500}$ and at a given redshift $z$ yield, in general, different lensing convergences: they have different shapes, being in general triaxial, and different large scale structure correlated and uncorrelated with them, which, in projection along the line of sight, also contributes to the convergence. Since $p_{\mathrm{obs}}$ is linear in the lensing convergence, this variability in the lensing convergence translates directly into variability in $p_{\mathrm{obs}}$.

We can write $p_{\mathrm{obs}}$ at a given point in true cluster parameter space, $M_{500}$--$z$, as a sum of three random variables (as in Eq.~\ref{eq:intro1}, repeated here for convenience),
\begin{equation}
p_{\mathrm{obs}} = p_{\textrm{c}} + \Delta p_{\textrm{u}} + \Delta p_{\textrm{n}}\, .
\end{equation}
Here, $\Delta p_{\textrm{n}}$ denotes the contribution to $p_{\mathrm{obs}}$ coming from the lensing reconstruction noise. It is the random variable associated with the response of our matched filter at the given true cluster parameters to the reconstruction noise $n$ (see Eq.~\ref{eq:recon}). Next, $\Delta p_{\textrm{u}}$ denotes the contribution coming from LSS uncorrelated with the cluster (hereafter and for concision, `uncorrelated LSS'). It is the random variable associated with the response of our matched filter at the given true cluster parameters to our random convergence maps with the considered integration length $\chi_l$. Finally, $p_{\textrm{c}}$ denotes the contribution to $p_{\mathrm{obs}}$ arising from the cluster itself (that is, from cluster morphology and orientation) \emph{and} from LSS along the line of sight correlated with the cluster (hereafter, `correlated LSS'). It is defined as the variable that arises from subtracting $\Delta p_{\textrm{u}}$ from $p$, the variable that results after averaging $p_{\mathrm{obs}}$ over $\Delta p_{\textrm{n}}$. In practice, in our deconvolution approach we determine the distribution of $p_c$ by deconvolving the distributions of $p$ and $\Delta p_{\textrm{u}}$ estimated from our mock data sets.

If some real $p_{\mathrm{obs}}$ measurements are to be used in a cosmological analysis (e.g., in order to determine the mass scale of the cluster sample), the conditional probability density function (pdf) followed by $p_{\mathrm{obs}}$ at the true cluster parameters, $P(p_{\mathrm{obs}}|M_{500},z)$, needs to be determined. In this paper we propose two different approaches in order to characterise this conditional pdf. In the first approach, which we refer to as the \textit{deconvolution} approach and which is developed in Section \ref{subsec:a}, we treat the scatter arising from reconstruction noise ($\Delta p_{\textrm{n}}$) and from uncorrelated LSS ($\Delta p_{\textrm{u}}$) as observational noise, and we then characterise the signal arising only from the cluster itself and from correlated LSS ($p_{\textrm{c}}$) with our mock observations. This was the approach underlying the hierarchical model of \citet{Zubeldia2019}; here we aim to justify the choices of priors on the CMB lensing bias and intrinsic scatter parameters of that work, $1-b_{\mathrm{CMBlens}}$ and $\sigma_{\mathrm{CMBlens}}$, respectively. In the second approach, which we refer to as the \textit{extrapolation} approach and which is studied in Section \ref{subsec:b}, we think of observational noise as being solely due to reconstruction noise ($\Delta p_{\textrm{n}}$), and then characterise the signal due to the cluster itself and to both correlated and uncorrelated LSS $(p_{\textrm{c}} + \Delta p_{\textrm{u}})$, extrapolating the results to the full line of sight from $z=0$ to last scattering. %We argue that the latter is a better approach, as it models the scatter arising from the cluster and from LSS better, albeit at the expense of some small loss of signal-to-noise.

\subsection{Deconvolution approach}\label{subsec:a}

\subsubsection{Method}\label{subsubsec:amethod}

In this approach, our matched filter has both the reconstruction noise and the lensing convergence power spectra in the inverse-variance weighting (see Eq.~\ref{filtval}). Let us first think of a hypothetical set of simulated observations in which the variation due to lensing by uncorrelated LSS is present from $z=0$ back to last scattering, and not just within a box of a given length along the line of sight, and where reconstruction noise is also present. In order to study $P(p_{\mathrm{obs}}|M_{500},z;\chi_{\star})$, we divide it into two layers\footnote{Here, $\chi_{\star}$ denotes that the integral along the line of sight is performed from $\chi = 0$ to $\chi = \chi_{\star}$ (i.e., to last scattering); a semicolon is used to stress the fact that, unlike the other two conditioning variables, $\chi_{\star}$ is not a cluster-related variable.}. 

The first layer is $P(p_{\mathrm{obs}}|p_{\mathrm{c}},M_{500},z;\chi_{\star})$, which is approximately a Gaussian distribution centred on $p_{\mathrm{c}}$ with unit standard deviation. Indeed, first, both $\Delta p_{\textrm{u}}$ and $\Delta p_{\textrm{n}}$ have zero expected values. We empirically check that $\Delta p_{\textrm{u}}$ has an expected value consistent with zero as a function of $M_{500}$ with our random data sets, which can be thought of as realisations of $\Delta p_{\textrm{u}}$ at a given integration length (see below). In addition, reconstruction noise is approximately Gaussian, and therefore so is $\Delta p_{\textrm{n}}$, which is linear in it. As we show below, for our experimental specifications, the standard deviation of $\Delta p_{\textrm{n}}$, $\sigma_{\Delta p_{\textrm{n}}}$, is significantly larger than the standard deviation of $\Delta p_{\textrm{u}}$, $\sigma_{\Delta p_{\textrm{u}}}$, across the mass range considered, and so the variance of $\Delta p_{\text{u}}+\Delta p_{\text{n}}$ is dominated by the reconstruction noise. For modest levels of non-Gaussianity of $\Delta p_{\text{u}}$ -- expected given the long integration length $\chi_\star$, the relatively high redshift of the lenses involved, and the effective low-pass filtering due to the inverse-variance weighting in the matched filter -- we therefore expect the distribution of $\Delta p_{\text{u}}+\Delta p_{\text{n}}$ to be close to Gaussian. We can calculate the variance of $\Delta p_{\text{u}}$ by taking the second moment of Eq. (\ref{pdef}) to find
\begin{equation}
\sigma_{\Delta p_{\textrm{u}}}^2 = \left[\int \frac{d^2 \bmath{L}}{2 \upi} \frac{|\kappa_{\textnormal{t}} (\bmath{L})|^2}{N_{\kappa}(\bmath{L})} \right]^{-1} \int \frac{d^2 \bmath{L}}{2 \upi} \frac{C^{\kappa \kappa}_L}{N_{\kappa}(\bmath{L})} \, ,
\end{equation}
where $C^{\kappa \kappa}_L$ is the lensing convergence power spectrum, and where, recall, $N_{\kappa}(\bmath{L}) = N^{\textnormal{(0)}}(L)+C_L^{\kappa \kappa}$. For $z=0.23$ we find $\sigma_{\Delta p_{\textrm{u}}}/\sigma_{\Delta p_{\textrm{n}}}=0.19$ for $M_{500}=2\times 10^{14} M_{\odot}$, and $\sigma_{\Delta p_{\textrm{u}}}/\sigma_{\Delta p_{\textrm{n}}}=0.22$ for $M_{500}=10^{15} M_{\odot}$; for $z=0.52$ we find $\sigma_{\Delta p_{\textrm{u}}}/\sigma_{\Delta p_{\textrm{n}}}=0.15$ for $M_{500}=2\times 10^{14} M_{\odot}$, and $\sigma_{\Delta p_{\textrm{u}}}/\sigma_{\Delta p_{\textrm{n}}}=0.18$ for $M_{500}=10^{15} M_{\odot}$.
%Since standard deviations add in quadrature, we can safely assume $\Delta p_{\textrm{u}} + \Delta p_{\textrm{n}}$ to be Gaussian as well.
Finally, our matched-filter inverse-variance weighting guarantees that $\Delta p_{\textrm{u}} + \Delta p_{\textrm{n}}$ has unit standard deviation and hence so does the distribution $P(p_{\mathrm{obs}}|p_{\mathrm{c}},M_{500},z;\chi_{\star})$.
In the context of this \textit{deconvolution} approach, we refer to $P(p_{\mathrm{obs}}|p_{\mathrm{c}},M_{500},z;\chi_{\star})$ as \textit{observational} scatter.

The other layer in which we decompose the scatter of our observable is $P(p_{\mathrm{c}}|M_{500},z;\chi_{\star})$. Within the context of this approach, we refer to this variability as \textit{intrinsic} scatter. We study this conditional distribution with our mock observations. First, we assume that $P(p_{\mathrm{c}}|M_{500},z;\chi_l = 400\,\mathrm{Mpc})$ is, to a good approximation, equal to $P(p_{\mathrm{c}}|M_{500},z;\chi_{\star})$; we remind that $\chi_l = 400\,\mathrm{Mpc}$ is the largest integration length that we have considered. Indeed, beyond $200$\,Mpc from the cluster centre, the contribution from correlated LSS to the convergence should be negligible (see Section \ref{subsec:b} for a quantitative discussion). Thus, $P(p_{\mathrm{c}}|M_{500},z;\chi_l = 400\,\mathrm{Mpc})$ is what needs to be determined. In the following, we will denote this pdf simply with $P(p_{\mathrm{c}}|M_{500},z)$.

\begin{figure}
\centering
\includegraphics[width=0.5\textwidth]{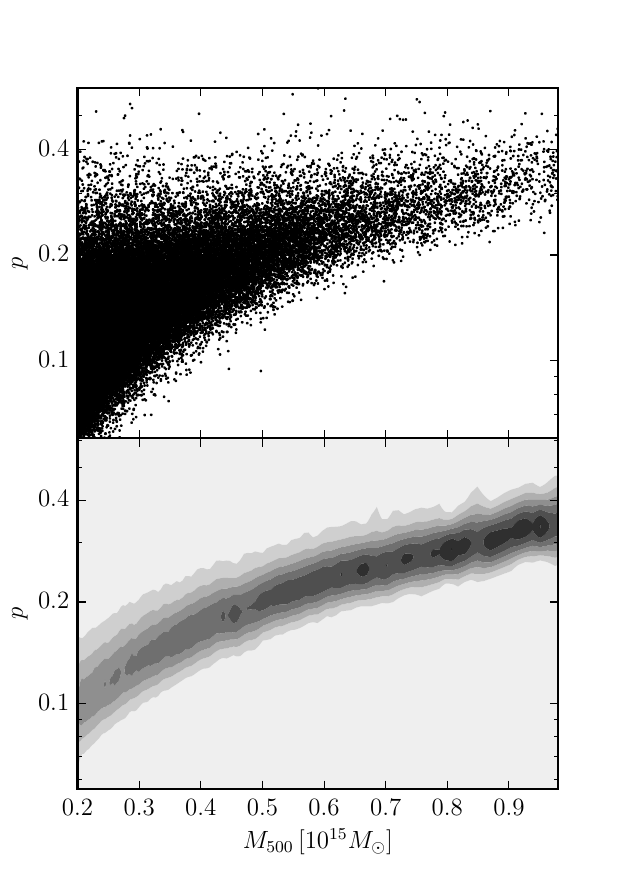}
\caption{Upper panel: scatter plot of $(p,M_{500})$ for $\chi_l=400$\,Mpc and $z = 0.23$. Lower panel: estimate of the conditional pdf of $p$, $P(p|M_{500},z;\chi_l)$, as a function of $M_{500}$ for the same case as the upper panel, as obtained with the kernel density estimation package fastKDE.}
\label{fig:kde_scatter}
\end{figure}

Unfortunately, our mock observations do not provide us with samples from $P(p_{\mathrm{c}}|M_{500},z)$. However, our cluster data set $A$ can be thought of as consisting of samples from $P(p,M_{500}|z;\chi_l)$, where $p = p_{\mathrm{c}} + \Delta p_{\textrm{u}}$, and where $z$ can be either of our two redshifts ($z=0.23$ and $z=52$), and $\chi_l$ either of our seven integration lengths (see Section \ref{sec:sims}). These observations for $\chi_l = 400$\,Mpc and $z = 0.23$ are shown, for illustrative purposes, in the upper panel of Fig.~\ref{fig:kde_scatter}. In addition, our random data set $A$ can be thought of as consisting of samples from $P(\Delta p_{\mathrm{u}}|M_{500},z;\chi_l)$ at the considered redshifts $z$, integration lengths $\chi_l$, and true masses $M_{500}$ (see Section \ref{sec:obs}). Our route to obtain $P(p_{\mathrm{c}}|M_{500},z)$ is first to determine $P(p|M_{500},z;\chi_l)$ as a function of $M_{500}$ as slices in $M_{500}$ of $P(p,M_{500}|z;\chi_l)$, and then, since $ = p_{\textrm{c}} + \Delta p_{\textrm{u}}$, to (formally) deconvolve it with $P(\Delta p_{\mathrm{u}}|M_{500},z;\chi_l)$. We describe this procedure, which takes its name from the deconvolution step, in detail in the rest of this section.

We first group our mock measurements of $p$ for $\chi_l=400$\,Mpc at $z=0.23$ into 90 subsets by binning their corresponding values of $M_{500}$ into 90 equally-spaced bins between $2 \times 10^{14} M_{\odot}$ and $10^{15} M_{\odot}$. For $z=0.52$ we apply a similar binning but then combine the 33 bins of the high-mass end into groups of three, yielding a total of 68 bins. We do this in order to compensate for the very small number of clusters per original bin at the high-mass end for $z=0.52$, which in some cases are below 10. We then assume that the distribution followed by $p$ in each of these bins is $P(p|M_{500},z;\chi_l = 400\,\mathrm{Mpc})$, with $M_{500}$ equal to the central value of the bin. This is certainly true in the limit in which the bins are infinitesimal, and should be a good approximation if they are sufficiently small. For illustrative purposes, an estimate of $P(p|M_{500},z;\chi_l = 400\,\mathrm{Mpc})$ as a function of $M_{500}$ can be seen in the lower panel of Fig.~\ref{fig:kde_scatter} for $z=0.23$ (the same case as the upper panel). It was produced with a kernel density estimation method, fastKDE \citep{OBrien2014,Obrien2016}; we note that this illustrative estimate is done with no previous binning.

%; numerical evidence that this is a good approximation in our case is provided in Appendix \ref{sec:appendix}

We treat each of the $M_{500}$ bins independently. Let us consider one of our snapshot redshifts $z$ and one of our $M_{500}$ bins. We can write $P(p|M_{500},z;\chi_l = 400\,\mathrm{Mpc})$ as
\begin{multline}
P(p|M_{500},z;\chi_l = 400\,\mathrm{Mpc}) =  \\ \int_0^{\infty} P(p|p_{\mathrm{c}},M_{500},z;\chi_l = 400\,\mathrm{Mpc}) P(p_{\mathrm{c}}|M_{500},z) dp_{\mathrm{c}} \, ,
\label{likelihood}
\end{multline}
where $P(p|p_{\mathrm{c}},M_{500},z;\chi_l = 400\,\mathrm{Mpc})$ is equal to $P(\Delta p_{\mathrm{u}}|M_{500},z;\chi_l = 400\,\mathrm{Mpc})$ evaluated at $\Delta p_{\text{u}}=p-p_{\mathrm{c}}$, and where $P(p_{\mathrm{c}}|M_{500},z)$ is the distribution of intrinsic scatter that we want to characterise (note that we have dropped the dependence on $\chi_l$ following the convention introduced above, as this pdf for $\chi_l=400$\,Mpc ought to be very close to that for $\chi_l = \chi_{\star}$). That is, as anticipated, the pdf of $p$ can be seen as a convolution of the pdf of $p_{\mathrm{c}}$ with the pdf of $\Delta p_{\mathrm{u}}$. We then assume that $P(p_{\mathrm{c}}|M_{500},z)$ is approximately log-normal. Log-normality is a common assumption for the intrinsic scatter of cluster observables, even if evidence for, e.g., some skewness has been found in some studies (see, e.g., \citealt{Becker2011}). In particular, we model $P(\ln p_{\mathrm{c}}|M_{500},z)$\footnote{Here and throughout, $\ln$ refers to the natural logarithm.} with an Edgeworth series, truncating it after the first three expansion terms.
That is, we assume that $P(\ln p_{\mathrm{c}}|M_{500},z)$ can be written as \citep{Wallace1958}
\begin{equation}\label{eq:edgeworth}
P(\ln p_{\mathrm{c}}|M_{500},z) = \frac{1}{\sigma_{\mathrm{c}}} \left(\phi - \frac{\lambda_{3,\mathrm{c}}}{6} \phi^{(3)} + \frac{\lambda_{\mathrm{4,c}}}{24} \phi^{(4)} + \frac{\lambda_{3,\mathrm{c}}^2}{72} \phi^{(6)}\right) \, .
\end{equation}
Here, $\phi$ is the standard normal distribution evaluated for $\left(\ln p_{\mathrm{c}} - \mu_{\mathrm{c}} \right) / \sigma_{\mathrm{c}}$, and $\phi^{(n)}$ is the $n$-th derivative of the standard normal distribution, also evaluated for $\left(\ln p_{\mathrm{c}} - \mu_{\mathrm{c}} \right) / \sigma_{\mathrm{c}}$. The quantities $\mu_{\mathrm{c}}$, $\sigma_{\mathrm{c}}$, $\lambda_{3,\mathrm{c}}$, and $\lambda_{4,\mathrm{c}}$ are, respectively, the mean, standard deviation, skewness, and kurtosis of $\ln p_{\mathrm{c}}$. The skewness and kurtosis, in particular, are defined in terms of the cumulants $k_n$ of $\ln p_{\mathrm{c}}$ as $\lambda_{3,\mathrm{c}} = k_3/\sigma_{\mathrm{c}}^3$ and $\lambda_{4,\mathrm{c}} = k_4/\sigma_{\mathrm{c}}^4$, respectively. As can be seen from Eq.~(\ref{eq:edgeworth}), a Gaussian distribution has $\lambda_{3,\mathrm{c}}=\lambda_{4,\mathrm{c}}=0$ (and all higher-order cumulants vanish also). We note that $\lambda_{4,\mathrm{c}}$ is sometimes referred to as excess kurtosis. We refer to this model of the intrinsic scatter as the log-Edgeworth model of the intrinsic scatter.

Assuming this model, $P(p_{\mathrm{c}}|M_{500},z)$ is characterised by four parameters, which we choose to be $\beta_{\mathrm{c}}$, $\sigma_{\mathrm{c}}$, $\lambda_{3,\mathrm{c}}$, and $\lambda_{4,\mathrm{c}}$. Here, $\beta_{\mathrm{c}}$ is a lensing mass bias parameter that is used in substitution of $\mu_{\mathrm{c}}$ and is defined as follows. As in \citet{Zubeldia2019}, we introduce the model mean signal-to-noise at true cluster parameters $M_{500}$ and $z$, $\bar{p}(\beta_{\mathrm{c}}M_{500},z)$, as
\begin{equation}\label{mfsnr}
\bar{p}\left( \beta_{\mathrm{c}} M_{500}, z\right) \equiv \frac{\kappa_0(\beta_{\mathrm{c}}M_{500},z)}{\sigma_{\kappa_0}(\beta_{\mathrm{c}}M_{500},z)} =
\frac{\beta_{\mathrm{c}}M_{500}}{\sigma_{M_{500}}\left(\beta_{\mathrm{c}} M_{500},z\right)}\, ,
\end{equation}
where $\kappa_0(\beta_{\mathrm{c}}M_{500},z)$ is the convergence of the truncated NFW model at the scale radius for a cluster of mass $\beta_{\mathrm{c}} M_{500}$, $\sigma_{\kappa_0}(\beta_{\mathrm{c}} M_{500},z)$ is the noise for the matched filter in this case, and 
$\sigma_{M_{500}}\left(\beta_{\mathrm{c}} M_{500},z\right)$ is the noise of the $\hat{M}_{500}$ matched filter estimator, given by Eq. (26) of \citet{Zubeldia2019}. Following \citet{Zubeldia2019}, the lensing mass bias parameter $\beta_{\mathrm{c}}$ is determined by demanding that $\ln \bar{p}(\beta_{\mathrm{c}}M_{500},z)$ is the mean of $\ln p_{\mathrm{c}}$ for true clusters of mass $M_{500}$ at redshift $z$, i.e., $\ln \bar{p}(\beta_{\mathrm{c}}M_{500},z) = \mu_{\mathrm{c}} $. As we discuss further below, deviations of $\beta_{\mathrm{c}}$ from unity arise both from mismatch between the true mean cluster (plus correlated LSS) convergence and the truncated NFW model at the same mass, and from intrinsic scatter. Although the model mean cluster signal-to-noise is not the expected value of any of our random variables, for small $\sigma_{\mathrm{c}}$ it is approximately equal to the expected value of $p_{\mathrm{c}}$, in which case the bias is determined by profile mismatch.
We note that in \citet{Zubeldia2019} $\beta_{\mathrm{c}}$ is denoted with $1 - b_{\mathrm{CMBlens}}$ and is assumed to be independent of mass and redshift.

\begin{figure}
\centering
\includegraphics[width=0.5\textwidth]{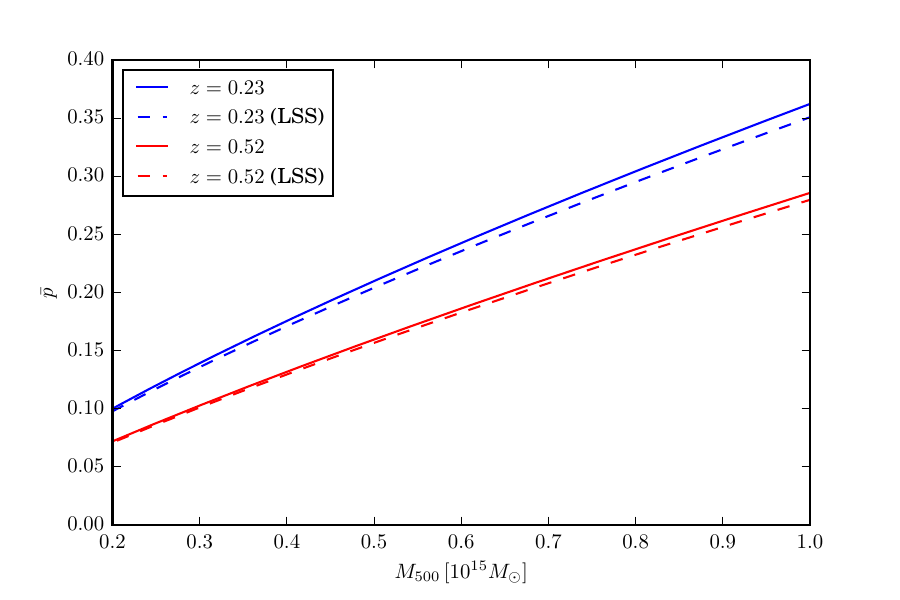}
\caption{Model mean signal-to-noise, $\bar{p}\left( \beta_{\mathrm{c}} M_{500}, z\right)$, as defined in Eq. (\ref{mfsnr}), as a function of $M_{500}$ for $z=0.23$ (blue curves) and $z=0.52$ (red curves) and for our two choices of matched filter inverse-variance weighting: one including only the reconstruction noise power spectrum (solid curves); and the other one also including the lensing convergence power spectrum (dashed curves). All the curves are computed using Eq.~(\ref{mfsnr}) assuming $\beta_{\mathrm{c}}=1$.}
\label{fig:mean_snr}
\end{figure}

The mean signal-to-noise $\bar{p}$, as defined in Eq. (\ref{mfsnr}), is shown in Fig.~\ref{fig:mean_snr} for our reference \textit{Planck}-like experiment as a function of $M_{500}$ for the two redshifts and the two matched filter inverse-variance weightings considered, one including the contribution from uncorrelated LSS (dashed lines), and the other one consisting only of reconstruction noise (solid lines). We remind that the former weighting is the one considered in this deconvolution approach; the other choice is also shown for comparison. It can be seen that including LSS in the inverse-variance weighting reduces the mean signal-to-noise at any given mass, as the observations are understood as being noisier. This decrement, however, is small, since reconstruction noise dominates over the variation due to uncorrelated LSS in our reference experiment. We note that $\bar{p}$ in Fig.~\ref{fig:mean_snr} is computed assuming $\beta_{\mathrm{c}}=1$ in Eq. (\ref{mfsnr}); $\bar{p}$ can be evaluated at any biased mass $\beta_{\mathrm{c}} M_{500}$ through simple interpolation.

Finally, we determine the remaining factor of the integrand of Eq. (\ref{likelihood}), $P(\Delta p_{\mathrm{u}}|M_{500},z;\chi_l = 400\,\mathrm{Mpc})$, with our random data set $A$. Recall, this data set consists of $10^5$ mock measurements of $\Delta p_{\mathrm{u}}$ for each mass bin, redshift, and integration length considered. Specifically, for each of our $M_{500}$ bins we estimate $P(\Delta p_{\mathrm{u}}|M_{500},z;\chi_l = 400\,\mathrm{Mpc})$ with the  corresponding $10^5$ mock measurements for $\chi_l=400$\,Mpc in a non-parametric way using the kernel density estimation library fastKDE \citep{OBrien2014,Obrien2016}.

Assuming our log-Edgeworth model for the intrinsic scatter and with the scatter due to uncorrelated LSS determined with kernel density estimation, Eq. (\ref{likelihood}) can be thought of as defining the likelihood of our mock measurements of $p$ for $\chi_l =400$\,Mpc for the four parameters, $\beta_{\mathrm{c}}$, $\sigma_{\mathrm{c}}$, $\lambda_{3,\mathrm{c}}$, and $\lambda_{4,\mathrm{c}}$. Adopting wide flat priors for all four parameters, we explore the corresponding posterior distributions across our $M_{500}$ bins and for our two snapshot redshifts with the \texttt{emcee} package, which performs affine-invariant MCMC sampling\footnote{\url{http://dfm.io/emcee/current/}}, generating a total of $10^5$ samples for each bin. In addition, we consider a second case in which we take the intrinsic scatter $P( p_{\mathrm{c}}|M_{500},z)$ to be log-normal. This model is the particular case of our log-Edgeworth model in which $\lambda_{\mathrm{3,c}}=\lambda_{\mathrm{4,c}} = 0$. In this case, the likelihood has only two parameters, $\beta_{\mathrm{c}}$ and $\sigma_{\mathrm{c}}$. Similarly adopting wide flat priors on the parameters, we explore the corresponding posteriors with the \texttt{emcee} package, also generating $10^5$ samples for each bin.

\subsubsection{Results and discussion}\label{subsec:resultsdisc}

%Figure full. Consistency, robustness, trends. Log-normal approximately good approximation for large masses, but fails at lower masses.

\begin{figure*}
\centering
\includegraphics[width=0.85\textwidth]{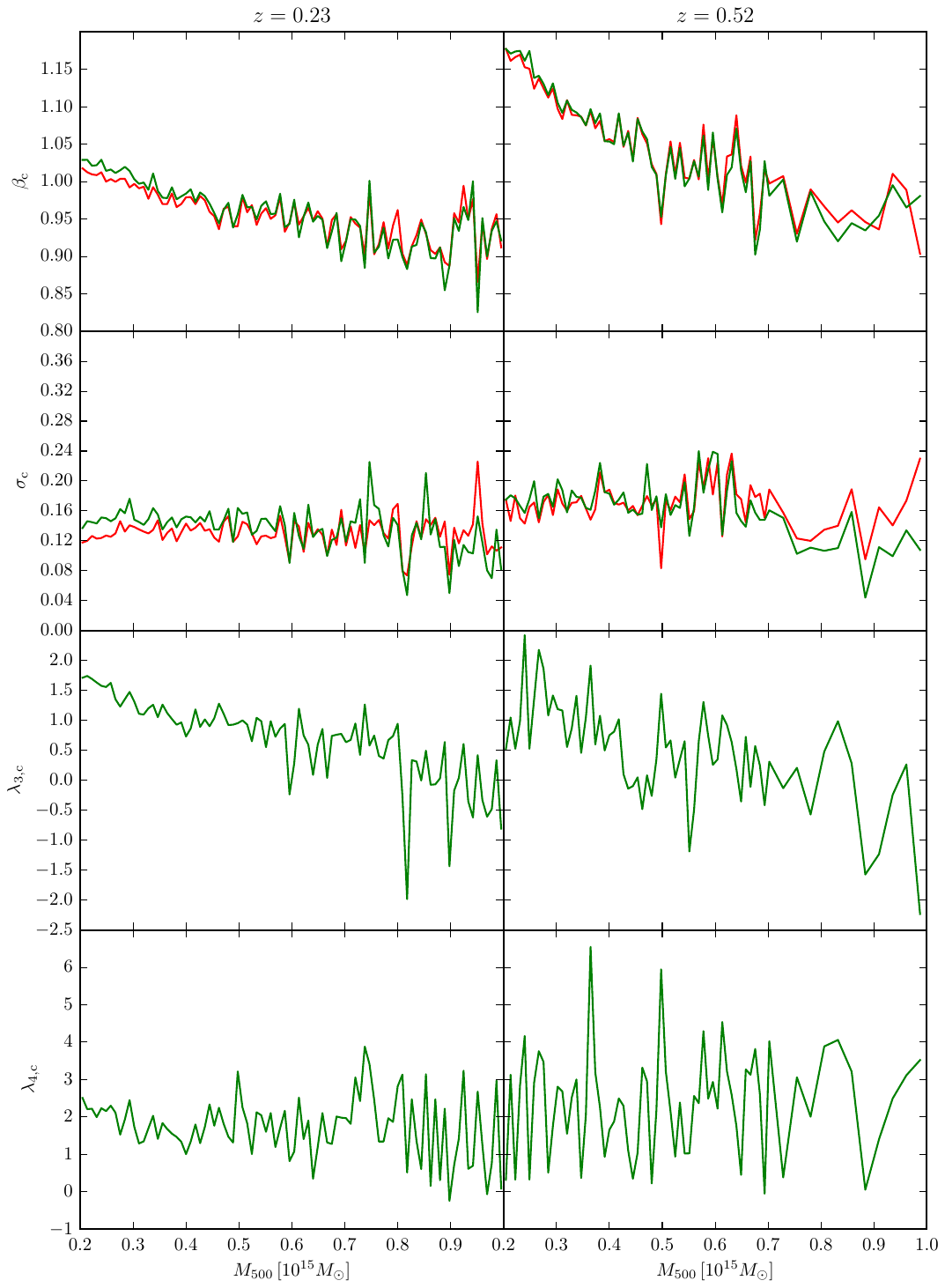}
\caption{Measured (median) values of $\beta_{\mathrm{c}}$ (lensing mass bias), $\sigma_{\mathrm{c}}$ (scatter), $\lambda_{3,\mathrm{c}}$ (skewness), and $\lambda_{4,\mathrm{c}}$ (kurtosis) as a function of $M_{500}$ for $z=0.23$ (left) and $z=0.52$ (right) obtained from the cluster and random data sets $A$ following our deconvolution approach (see Section \ref{subsec:a}). The parameter values obtained assuming our log-Edgeworth model are shown in green, whereas those obtained assuming log-normality are shown in red. We note that the noise visibly increases with mass, as the number of clusters per bin decreases. For $z=0.52$, however, the noise decreases at the high-mass end due to the use of wider bins.}
\label{fig:edgeworth_full}
\end{figure*}

\begin{figure*}
\centering
\includegraphics[width=0.8\textwidth]{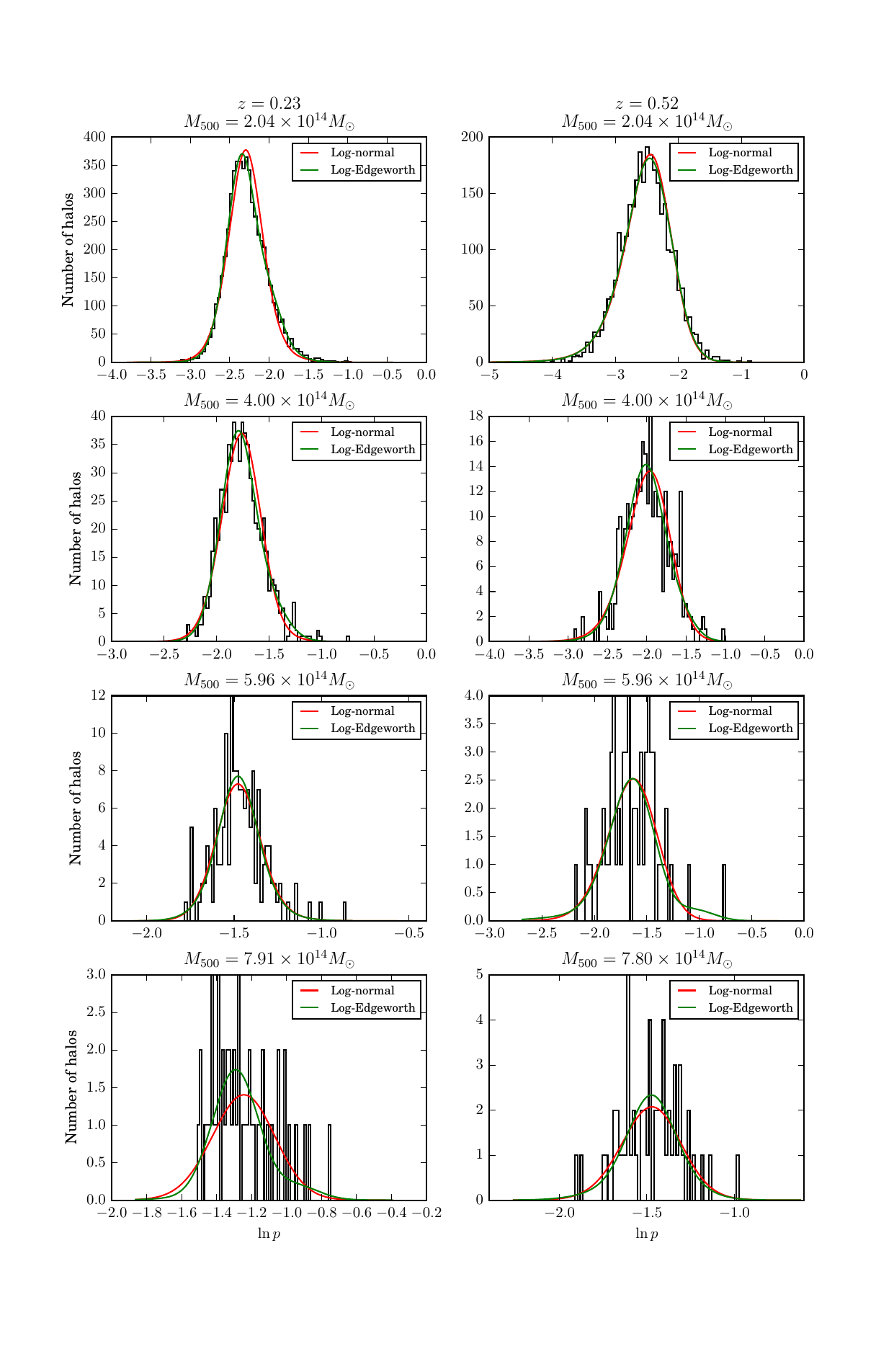}
\caption{Histograms of the values of $\ln p$ from the cluster data set $A$ for $\chi_l=400\textrm{\,Mpc}$ for several selected $M_{500}$ bins and at the two considered snapshot redshifts, $z=0.23$ (left panels), and $z=0.52$ (right panels). In addition, corresponding predictions from the log-Edgeworth model of the intrinsic scatter for the median parameter values as obtained from the MCMC samples are shown in green, and analogous predictions from the log-normal model of the intrinsic scatter (see Section \ref{subsubsec:amethod}) are shown in red.}
\label{fig:histograms}
\end{figure*}

Figure \ref{fig:edgeworth_full} shows the measured values of $\beta_{\mathrm{c}}$, $\sigma_{\mathrm{c}}$, $\lambda_{3,\mathrm{c}}$, and $\lambda_{4,\mathrm{c}}$ as a function of $M_{500}$ for our two snapshot redshifts that we obtain with our deconvolution approach. Specifically, for each $M_{500}$ bin the median value of each parameter as obtained from the MCMC samples is shown. The green curves are obtained assuming the log-Edgeworth model of the intrinsic scatter described in Section \ref{subsubsec:amethod}, whereas the red curves are obtained taking the intrinsic scatter to be log-normal. By construction, $\lambda_{3,\mathrm{c}}=\lambda_{4,\mathrm{c}}=0$ for the log-normal case; these constant zero lines are not shown for clarity. The curves are, as expected, noisy, due to the fact that there is only a finite number of clusters in each $M_{500}$ bin. In general, the noise visibly increases with mass, as the number of clusters per bin decreases. For $z=0.52$, however, the noise decreases at the high-mass end due to the use of wider bins.

In addition, Fig.~\ref{fig:histograms} shows the histograms of $\ln p$ for our simulated clusters (with
$\chi_l=400\textrm{\,Mpc}$) in several selected $M_{500}$ bins at both snapshot redshifts. The predictions of our two models, Eq.~\eqref{likelihood} with log-Edgeworth or log-normal intrinsic scatter, evaluated at the respective median parameter values obtained from our MCMC samples, are also shown as the green and red curves, respectively.

%DISCUSSION

%Detection of skewness and kurtosis

Several remarks can be made about Figs~\ref{fig:edgeworth_full} and \ref{fig:histograms}. First, in Fig.~\ref{fig:edgeworth_full} it can be seen that the log-Edgeworth model detects significant non-zero (positive) skewness, $\lambda_{3,\mathrm{c}}$, and kurtosis, $\lambda_{4,\mathrm{c}}$, across most of our mass bins and for both redshifts. At fixed redshift, $\lambda_{3,\mathrm{c}}$ is observed to decrease with mass, with no significant detection at the high mass end. On the other hand, $\lambda_{4,\mathrm{c}}$ is observed to depend less strongly on mass and to be non-zero at all masses. Neither of these parameters exhibit a strong redshift dependence. The bias, $\beta_{\mathrm{c}}$, and scatter, $\sigma_{\mathrm{c}}$, obtained assuming the log-Edgeworth model take values close to those obtained assuming log-normality. A small systematic difference is observed at lower masses, especially for $z=0.23$, which we attribute to the detection of non-zero skewness and kurtosis,
%$\lambda_{3,\mathrm{c}}$ and $\lambda_{4,\mathrm{c}}$, respectively,
since these two parameters have some degeneracy with $\beta_{\mathrm{c}}$ and $\sigma_{\mathrm{c}}$.

%Describe trends in bias and scatter. Can be improved

For both models, $\beta_{\mathrm{c}}$, is observed to depend on both mass and redshift. At fixed redshift, it decreases with mass. At fixed mass, it is observed consistently to take larger values at $z=0.52$ than at $z=0.23$, the difference being larger at smaller masses. The scatter, $\sigma_{\mathrm{c}}$, on the other hand, does not have a strong dependence on mass for either model. However, there does seem to be a small dependence with redshift; at a given mass, $\sigma_{\mathrm{c}}$ generally takes larger values at $z=0.52$.

%Interpretation of scatter

We recall that $\sigma_{\mathrm{c}}$ is the standard deviation of $\ln p_{\mathrm{c}}$, the logarithm of the contribution to the signal-to-noise associated with the cluster itself and to correlated LSS. It is therefore a measure of how large the fractional scatter on $p_{\mathrm{c}}$, $\sigma_{p_{\mathrm{c}}} / p_{\mathrm{c}}$, is, where $\sigma_{p_{\mathrm{c}}}$ is the standard deviation of $p_{\mathrm{c}}$. Thus, we find that the standard deviation of $p_{\mathrm{c}}$ due to the cluster itself and to correlated LSS increases with mass at roughly the same rate as $p_{\mathrm{c}}$ does. 

%Interpretation of bias

The bias, $\beta_{\mathrm{c}}$, on the other hand, has a less straightforward interpretation. It depends both on the mismatch between the convergence model used in the matched filter (in our case, a truncated NFW profile; see Section \ref{sec:obs}) and the true mean convergence at the given $M_{500}$ and $z$, and on the scatter (and higher moments) at that point in true cluster parameter space. This is a consequence of $\beta_{\mathrm{c}}$ being defined through the expected value of $\ln p_{\mathrm{c}}$, $\mu_{\mathrm{c}}$ (and not of, e.g., $p_{\mathrm{c}}$). We remind that $\mu_{\mathrm{c}} = \ln \bar{p}\left( \beta_{\mathrm{c}} M_{500}, z\right)$, where $\bar{p}\left( \beta_{\mathrm{c}} M_{500}, z\right)$ is the model mean signal-to-noise defined in Eq. (\ref{mfsnr}).
%which we remark is \emph{not} the expected value of any of our random variables.
In order understand this better, let us assume that the intrinsic scatter is log-normal. In this case, 
\begin{equation}\label{lognormalmeanc}
\ln \bar{p}\left( \beta_{\mathrm{c}} \right) = \ln \left\langle p_{\mathrm{c}} \right\rangle - \frac{\sigma_{\mathrm{c}}^2}{2} \, ,
\end{equation}
where $\left\langle p_{\mathrm{c}} \right\rangle$ is the expected value of $p_{\mathrm{c}}$, which only depends on the true mean convergence at the point in cluster parameter space considered, and which is equal to the true expected value of $p$, since uncorrelated large scale structure has zero expected value. Note that to simplify the notation, we have only made explicit the dependence of $\bar{p}$ on $\beta_{\mathrm{c}}$ in Eq.~\eqref{lognormalmeanc}.
Thus, in general, at fixed $M_{500}$ and $z$, $\beta_{\mathrm{c}}$ depends on both the true mean cluster convergence (through $\left\langle p_{\mathrm{c}} \right\rangle$) and the scatter, $\sigma_{\mathrm{c}}$. However, in our case $\sigma_{\mathrm{c}} \approx 0.15$, so $\sigma_{\mathrm{c}}^2/2 \approx 0.01$, and $\left\langle p_{\mathrm{c}} \right\rangle$ is typically $0.2$, so $\ln \left\langle p_{\mathrm{c}} \right\rangle \sim -1.6$. Therefore, $\ln \bar{p}\left( \beta_{\mathrm{c}} \right) \approx \ln \left\langle p_{\mathrm{c}} \right\rangle$, i.e., the bias, $\beta_{\mathrm{c}}$, essentially corrects for the mismatch between the assumed convergence model used in the matched filter and the true mean convergence. This is also true if we approximate the intrinsic scatter with our Edgeworth series; indeed, in this case we find
\begin{equation}\label{edgeworthmean}
\ln \bar{p}\left( \beta_{\mathrm{c}} \right) = \ln \left\langle p_{\mathrm{c}} \right\rangle - \frac{\sigma_{\mathrm{c}}^2}{2} - \ln\left( 1 + \frac{\lambda_{3,\mathrm{c}}}{6} \sigma_{\mathrm{c}}^3 + \frac{\lambda_{4,\mathrm{4}}}{24} \sigma_{\mathrm{c}}^4 + \frac{\lambda^2}{72} \sigma_{\mathrm{c}}^6 \right)\, .
\end{equation}
For $\sigma \approx 0.15$, $\lambda_{3,\mathrm{c}} \approx 1$, and $\lambda_{4,\mathrm{c}} \approx 2$, the new (logarithm) term is $\mathcal{O} (10^{-4})$, and therefore also negligible.  This also means that, if the scatter is small, the skewness and kurtosis also have little impact on the bias. We will see in Section \ref{subsec:choicetemplate} that using a model that better fits the true mean convergence yields $\beta_{\mathrm{c}} \approx 1$. In addition, in Section \ref{subsec:b}, where we consider the extrapolation approach, we will meet an example of the scatter not being negligible and the bias being sensitive to it as a result.

%Goodness of fit

Finally, in Fig.~\ref{fig:histograms} it can be seen that both models fit our mock data rather well across the mass bins and for both redshifts. Visually, the log-Edgeworth model (green curves) fits the mock data better than the log-normal model (red curves), as expected since the latter is a special case of the former and since statistically-significant skewness and kurtosis are detected throughout the mass bins. This is most apparent in the histograms corresponding to lower masses, which are less noisy as they contain a larger number of clusters (see, e.g., the upper-left panel). We note that rigorous analysis of the goodness of fit of each model and model comparison between the two models, which could be done by, e.g., comparing their Bayesian evidences as a function of mass and redshift, are beyond the scope of this paper.

\subsubsection{Sufficiency of log-normal approximation}\label{sufficiency}

\begin{figure}
\includegraphics[width=0.5\textwidth]{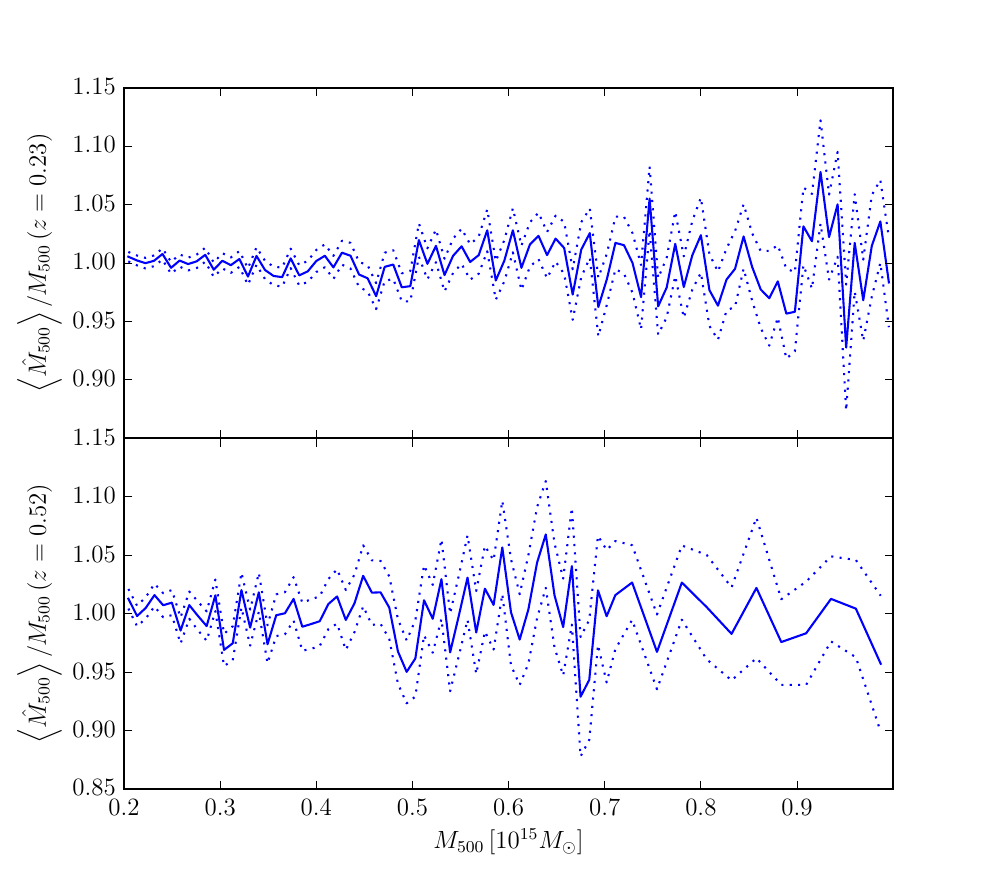}
\caption{Ratio of the expected value of the estimated mass, $\langle \hat{M}_{500}\rangle$, over the true mass, $M_{500}$, as a function of $M_{500}$ for $z=0.23$ (solid curve, upper panel) and $z=0.52$ (solid curve, lower panel), obtained as detailed in Section \ref{sufficiency}. The dotted curves depict the associated standard deviations. No evidence for a bias is seen.}
\label{fig:unbiased}
\end{figure}

In the previous section we presented evidence for the intrinsic scatter not being log-normal (see Fig.~\ref{fig:edgeworth_full}). A question that arises as a consequence is whether assuming log-normal intrinsic scatter is a good enough approximation. Here we show that for \textit{Planck} clusters log-normal intrinsic scatter is indeed a good approximation that yields unbiased estimates of $M_{500}$.

We show that assuming log-normal intrinsic scatter is adequate as follows. First, we fit a quadratic polynomial to our median values of $\beta_{\mathrm{c}}$ and $\sigma_{\mathrm{c}}$ obtained assuming log-normal intrinsic scatter (red curves in Fig.~\ref{fig:edgeworth_full}). We do this by minimising the mean square error, taking the noise associated to each data point (i.e., to each $M_{500}$ bin) to be inversely proportional to the square root of the number of halos within that bin. This produces smooth estimates of $\beta_{\mathrm{c}}$ and $\sigma_{\mathrm{c}}$ as a function of $M_{500}$ which, for each $M_{500}$ bin, are much less sensitive to the specific random fluctuation in that bin than the original data point is. Then, for each $M_{500}$ bin we use the likelihood that we used to determine $\beta_{\mathrm{c}}$ and $\sigma_{\mathrm{c}}$, defined in Eq. (\ref{likelihood}), assuming log-normal intrinsic scatter. This function, fed with our smooth estimates of $\beta_{\mathrm{c}}$ and $\sigma_{\mathrm{c}}$ (which we evaluate at the central mass of the bin) and with the corresponding mock measurements of $p$, can be seen as a posterior for $M_{500}$ (after taking the prior on $M_{500}$ to be flat and wide). Since it is a one-dimensional posterior, we explore it by evaluating it for a range of values of $M_{500}$ and we compute the expected value, which we denote with $\langle \hat{M}_{500} \rangle$, and the standard deviation, which we denote with $\sigma_{\hat{M}_{500}}$. Figure \ref{fig:unbiased} shows this mean estimated mass over the true mass $M_{500}$, $\langle \hat{M}_{500}\rangle / M_{500}$, as a function $M_{500}$ for our two snapshot redshifts, along with the corresponding standard deviation, $\sigma_{\hat{M}_{500}} / M_{500}$ (depicted with the dotted lines). No evidence for a bias is seen.

\subsubsection{Impact of the choice of cluster convergence model}\label{subsec:choicetemplate}

\begin{figure*}
\includegraphics[width=0.6\textwidth]{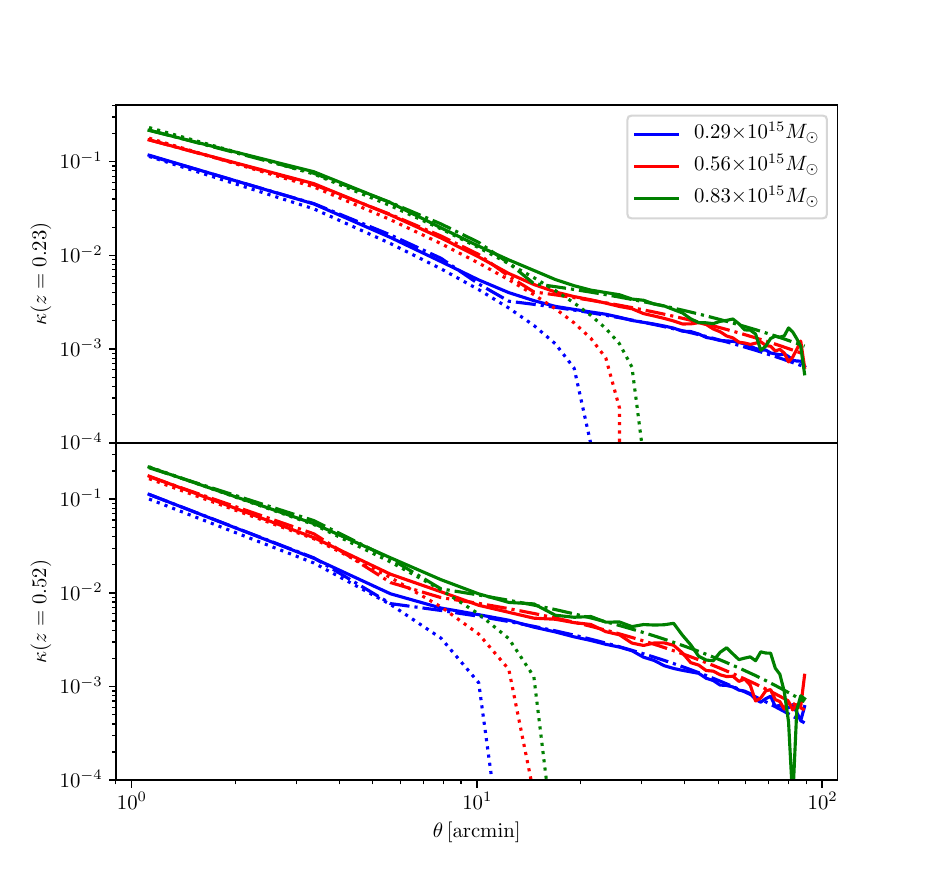}
\caption{Azimuthally-averaged convergence as a function of angular separation from the cluster centre, $\theta$, for three representative $M_{500}$ bins, each identified by a different colour, for $z=0.23$ (upper panel) and $z=0.52$ (lower panel). The solid curves show the empirical convergence as obtained from our convergence maps extracted from the simulation snapshots; the dotted curves show the prediction from our truncated NFW model; and the dash-dotted curves show the prediction from our fitted truncated NFW+2h model. The latter model visibly matches the mean simulation convergence better than the former one.}
\label{fig:annuli}
\centering
\end{figure*}

\begin{figure}
\includegraphics[width=0.4\textwidth]{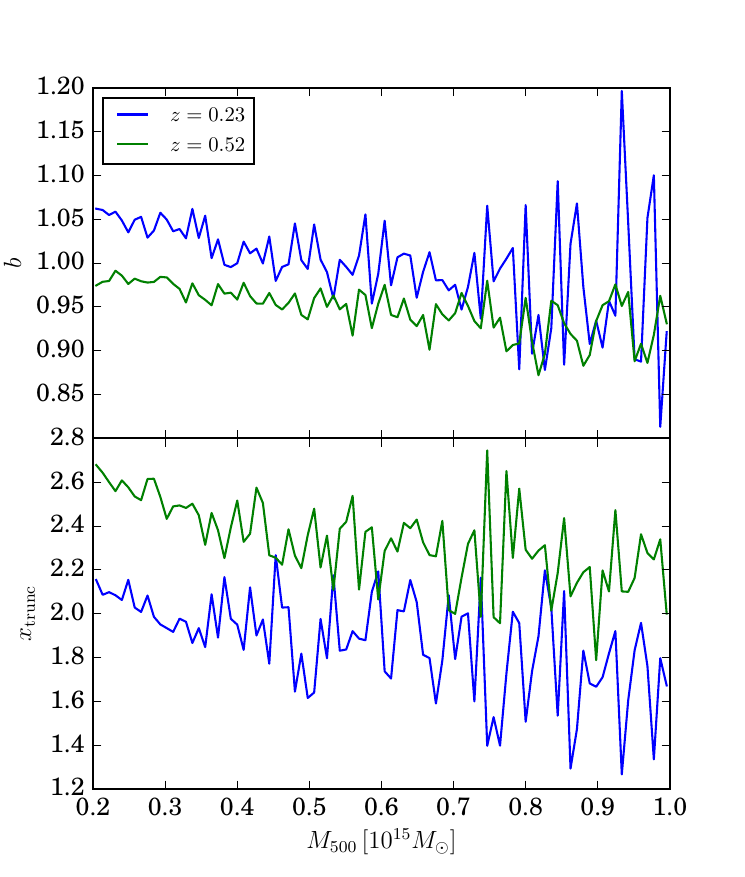}
\centering
\caption{Values of the two free parameters of the fitted truncated NFW+2h model, $b$ and $x_{\mathrm{trunc}}$, when fitted to the mean cluster convergence, as a function of $M_{500}$ for $z=0.23$ (blue curves) and $z=0.52$ (green curves).}
\label{fig:fittingprams}
\end{figure}

\begin{figure}
\includegraphics[width=0.5\textwidth]{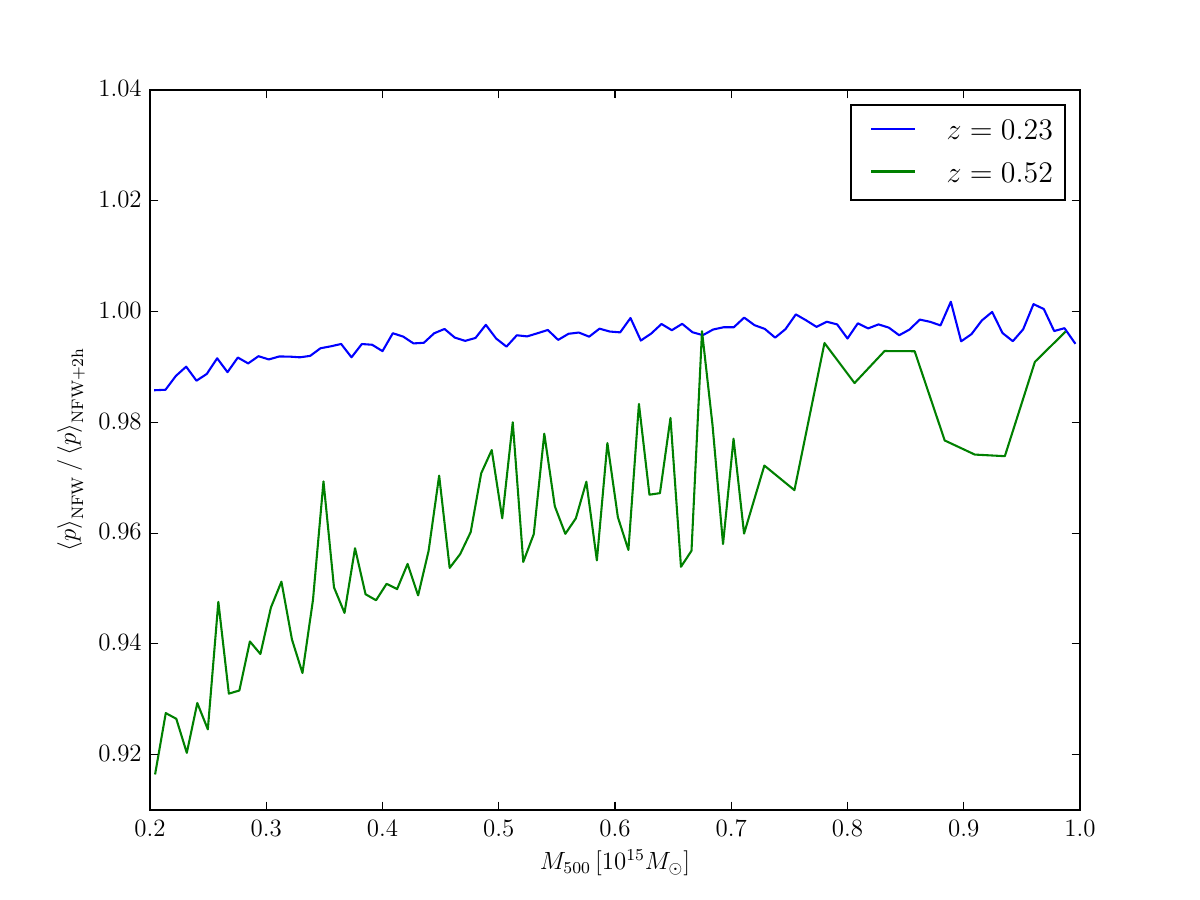}
\caption{Ratio between the expected value of $p_{\mathrm{c}}$, $\left\langle p_{\mathrm{c}} \right\rangle$, if the truncated NFW convergence model is used in the matched filter and the analogous quantity if the fitted truncated NFW+2h convergence model is used instead, as a function of $M_{500}$ for $z=0.23$ (blue curves) and $z=0.52$ (green curves), as obtained from our mock observations.}
\label{fig:ratio_mean}
\centering
\end{figure}

\begin{figure}
\includegraphics[width=0.5\textwidth]{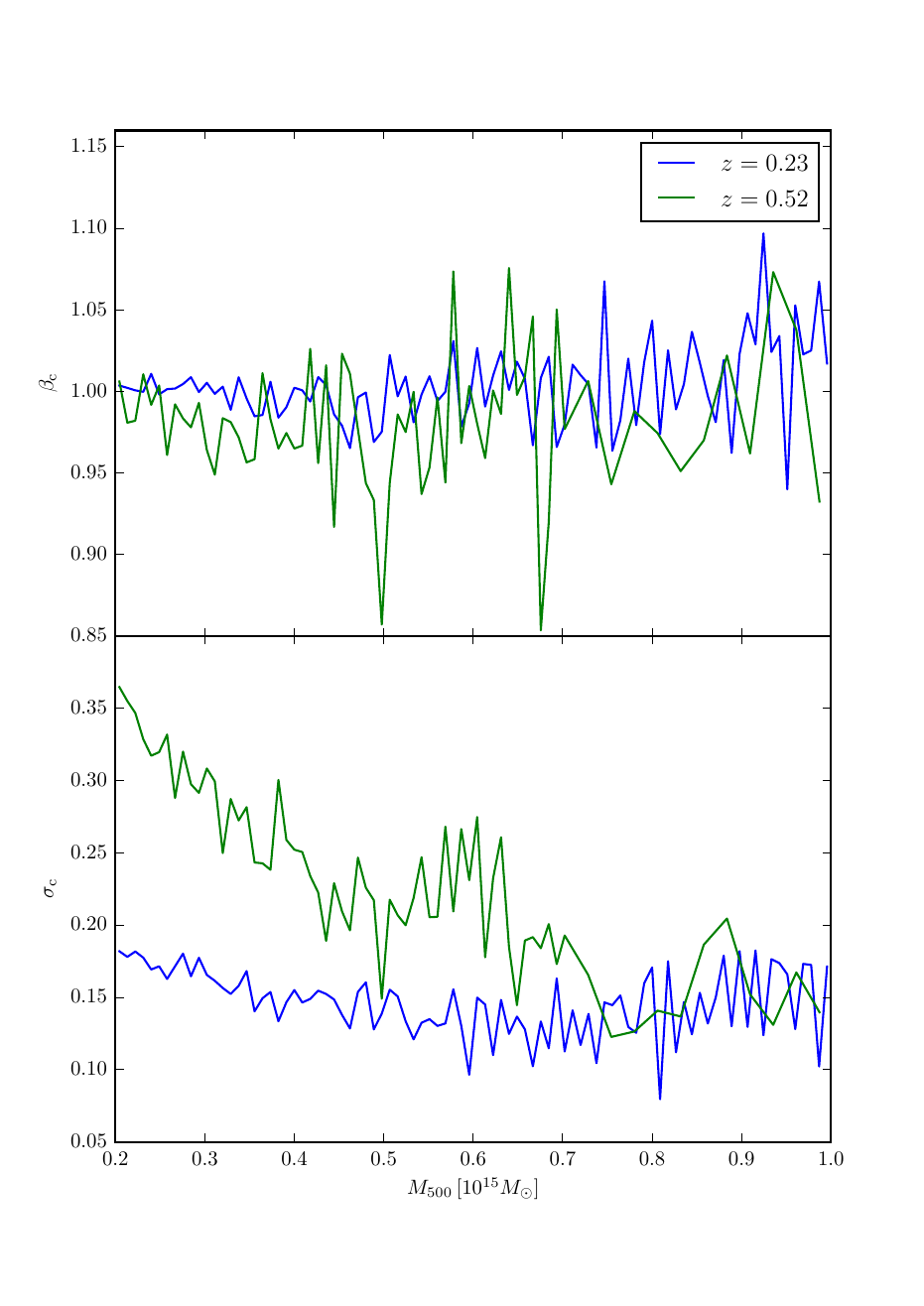}
\caption{Lensing mass bias, $\beta_{\mathrm{c}}$ (upper), and scatter, $\sigma_{\mathrm{c}}$ (lower), as a function of $M_{500}$ for $z=0.23$ (blue curves) and $z=0.52$ (green curves) obtained following the deconvolution approach using our fitted truncated NFW+2h convergence model in the matched filter.}
\label{fig:twohalo}
\end{figure}

As argued in Section \ref{subsec:resultsdisc}, at a given point in true cluster parameter ($M_{500}$--$z$) space, the mismatch between the lensing convergence model used in the matched filter and the true mean lensing convergence profile is the main contribution to the bias, $\beta_{\mathrm{c}}$. In particular, if the model matches the true mean lensing profile, then $\ln \bar{p} = \ln \left \langle p_{\mathrm{c}} \right \rangle$ and therefore $\beta_{\mathrm{c}} \approx 1$. In principle, any reasonable model can be used, and any mismatch between it and the true mean convergence profile will make $\beta_{\mathrm{c}} \neq 1$; its precise value can be determined as we have described in Sections \ref{subsubsec:amethod} and \ref{subsec:resultsdisc}. However, in a matched filter the signal-to-noise is optimal when the model matches exactly the true signal to be filtered; therefore, using a different template may yield suboptimal results. In this section we validate numerically that $\beta_{\mathrm{c}} \approx 1$ when the model fits the true mean convergence well enough by studying a new set of mock match-filtered observations produced with a more realistic convergence model and we investigate this potential issue of suboptimality.

The model that we use in the matched filter in order to produce the mock measurements studied in this paper is, as described in Section \ref{sec:obs}, an NFW profile with fixed concentration $c_{500} = 3$ (critical) and truncated at $R_{\mathrm{trunc}} = 5 R_{500}$. This was also the model used in \citet{Zubeldia2019}. Here we consider a more realistic cluster convergence model, which we write as
\begin{equation}
\kappa_{\mathrm{m}} (M_{500},z;b,x_{\mathrm{trunc}}) = \kappa_{\mathrm{1h}} (M_{500},z;b,x_{\mathrm{trunc}}) + \kappa_{\mathrm{2h}} (M_{500},z) \, .
\end{equation}
Here, $\kappa_{\mathrm{1h}} (M_{500},z;b,x_{\mathrm{trunc}})$ is the convergence of a truncated NFW profile with a concentration  $c_{500} = 2.6$, which is closer to the values reported by \citet{Diemer2015} for the mass and redshift range considered here (see the left panel of their Fig.~2), evaluated at a biased mass $b M_{500}$ and truncated at $R_{\mathrm{trunc}} = x_{\mathrm{trunc}} b^{1/3} R_{500}$. The second term in the model convergence, $\kappa_{\mathrm{2h}} (M_{500},z)$, is the two-halo term, which accounts for LSS correlated with the halo. We compute it following \citet{OguriH2011} and using the \citet{Tinker2010} model of the halo bias. We refer to the model $\kappa_{\mathrm{m}}$ as our fitted truncated NFW+2h model; for comparison purposes, in this section we refer to our original model as our truncated NFW model. For each $M_{500}$ bin, we determine $b$ and $x_{\mathrm{trunc}}$ by fitting $\kappa_{\mathrm{m}} (M_{500},z;b,x_{\mathrm{trunc}})$ to the corresponding mean convergence as obtained from our cluster convergence fields. More specifically, for each $M_{500}$ bin we obtain a mean convergence map by averaging over the convergence maps of all the clusters within it, and then we azimuthally average this mean convergence map over 40 annuli linearly spaced between the halo centre and an angular radius of $90$\,arcmin. We then fit for $b$ and $x_{\mathrm{trunc}}$ by minimising the mean square error between this binned convergence and the prediction from our model, $\kappa_{\mathrm{m}} (M_{500},z;b,x_{\mathrm{trunc}})$, taking the variance of each annulus to be inversely proportional to its area. The azimuthally-averaged lensing convergence profile as a function of angular separation from the halo centre measured form our cluster convergence fields can be seen in Fig.~\ref{fig:annuli} for three selected $M_{500}$ bins (solid curves) at our two snapshot redshifts, along with the corresponding predictions from our truncated NFW model (dotted curves) and our fitted truncated NFW+2h model (dash-dot curves). Visually, the latter model clearly fits the measured average convergence better. The fitted values of the two free parameters of this model, $b$ and $x_{\mathrm{trunc}}$, are shown in Figure \ref{fig:fittingprams}. The bias, $b$, deviates from $b=1$ (no bias) at the few-percent level, whereas the truncation parameter, $x_{\mathrm{trunc}}$, takes values around $2$, indicating that the two-halo term is already significant at $r \sim 2 R_{500}$. We recall that in our truncated NFW model we take $x_{\mathrm{trunc}} = 5$, which means that, through the use of a larger truncation radius, this model partially accounts for the two-halo term contribution within $r = 5 R_{500}$.

We then produce a new set of mock observations from our cluster and random convergence maps for $\chi_l=400$\,Mpc in exactly the same way that we produced our cluster and random data sets used in Section \ref{subsec:resultsdisc} (i.e., as described in Section~\ref{sec:obs}) but using our fitted truncated NFW+2h cluster convergence model in the matched filter instead of our truncated NFW model. The ratio between the expected value of $p_{\mathrm{c}}$, $\left\langle p_{\mathrm{c}} \right\rangle$ (which is equal to the expected value of $p$, since the contribution from uncorrelated LSS has zero expected value), for this fitted truncated NFW+2h model and the analogous quantity for our truncated NFW model is shown in Fig.~\ref{fig:ratio_mean} as a function of $M_{500}$ for our two snapshot redshifts. As expected, the fitted truncated NFW+2h model yields a higher mean signal-to-noise, since it fits the signal to be matched by the matched filter better; the difference is larger for $z=0.52$. Next, we apply our deconvolution approach, as described in \ref{subsubsec:amethod}, to these new mock observations in order to obtain the values of $\beta_{\mathrm{c}}$ and $\sigma_{\mathrm{c}}$ for this convergence model; we only consider log-normal scatter. These values are shown in Fig.~\ref{fig:twohalo} as a function of $M_{500}$ for our two snapshot redshifts. As anticipated in Section \ref{subsec:resultsdisc}, $\beta_{\mathrm{c}} \approx 1$ throughout our mass bins. At high masses, $\sigma_{\mathrm{c}}$ takes similar values to those that we obtain with our truncated NFW model (see Fig.~\ref{fig:edgeworth_full}), but at low masses it takes visibly higher values, especially for $z=0.52$. In fact, for this redshift, at low masses $\beta_{\mathrm{c}}$ is visibly smaller than $1$, which we attribute to the corresponding larger values of $\sigma_{\mathrm{c}}$ (as detailed in Section \ref{subsec:resultsdisc}, at fixed mean convergence, a larger value of $\sigma_{\mathrm{c}}$ yields a lower value of $\beta_{\mathrm{c}}$). We attribute this to the fact that it is more spatially extended, and therefore more sensitive to variations around the mean convergence due to correlated LSS.

%\AC{Is there an explanation we can give for the larger intrinsic scatter at lower masses for the fitted truncated NFW+2h model in the matched filter? Have you ever looked in real-space at the correctly-normalised filters for the signal-to-noise in the two cases (truncated NFW and fitted truncated NFW+2h), $\tilde{\kappa}(\theta)$, such that
%
%\begin{equation}
%p = \int d^2 \boldsymbol{\theta}\, \kappa(\theta) \tilde{\kappa}(\theta) \, ?
%\nolabel    
%\end{equation}
%
%Based on these, presumably one can make an argument about the fitted truncated NFW+2h case being more extended and picking up profile variations out to larger radii. (I recall you making such arguments when we discussed this in the past.)
%}

The fact that the mean signal-to-noise is larger for the fitted truncated NFW+2h model than it is for the truncated NFW model, but that the intrinsic scatter is also larger for the fitted truncated NFW+2h model at lower masses, poses the question of which model is more optimal, i.e., which one yields, after calibration, tighter constraints on $M_{500}$? We investigate this question by using new mock observations drawn as samples from our hierarchical model. Let us consider one of our two convergence models. First, for each redshift we fit a polynomial to the corresponding measurements of $\beta_{\mathrm{c}}$ and $\sigma_{\mathrm{c}}$ as a function of $M_{500}$ obtained assuming log-normality (the posterior median values; i.e., the values shown in Figs~\ref{fig:edgeworth_full} and \ref{fig:twohalo}). This yields a smooth estimate of $\beta_{\mathrm{c}}$ and $\sigma_{\mathrm{c}}$ as a function of $M_{500}$ for each redshift, which we take to be the true values of these parameters. This smoothing of $\beta_{\mathrm{c}}$ and $\sigma_{\mathrm{c}}$ is necessary for the subsequent posteriors of $M_{500}$ to be smooth (see below). Then, using these smooth estimates and Eq. (\ref{mfsnr}), we compute $\bar{p}$ at the central value of each of our $M_{500}$ bins. Next, for each bin we draw $10^4$ samples from a Gaussian centred at $\ln \bar{p}$ with standard deviation equal to the smooth estimate of $\sigma_{\mathrm{c}}$ at the bin centre. We exponentiate these samples, which yields a set of samples of $p_{\mathrm{c}}$ for each bin, and we finally add, to each of them, a sample from a Gaussian with zero mean and standard deviation unity. This procedure generates a set of mock observations of $p_{\mathrm{obs}}$ drawn according to our hierarchical model (with log-normal intrinsic scatter). We apply this procedure to both convergence models, using the same random seed for each model so that the random fluctuations are the same.

Next, we use these mock observations of $p_{\mathrm{obs}}$ for each convergence model in order to constrain, for each mass bin, its corresponding value of $M_{500}$. We do this by considering the likelihood associated with the hierarchical model used to generate these mock observations: fixing the values of $\beta_{\mathrm{c}}$ and $\sigma_{\mathrm{c}}$ to their (smooth) input values, this likelihood can be seen as a function of $M_{500}$. Assuming a wide flat prior, it yields a posterior for $M_{500}$ for each mass bin. This is very similar to the way in which we obtained a posterior for $M_{500}$ in Section \ref{sufficiency}, where we tested the impact of assuming log-normal intrinsic scatter; the only difference is that the pdf with which we convolve the intrinsic scatter pdf is the unit standard-deviation Gaussian that describes the full observational scatter, and not the pdf associated with the uncorrelated LSS within a given integration length, as is the case in Section \ref{sufficiency}. Each posterior is one-dimensional, so we explore them by evaluating them at a range of values of $M_{500}$ and then we compute their standard deviations, which we denote with $\sigma_{\hat{M}_{500}}^{\mathrm{NFW}}$ for the truncated NFW convergence model and with $\sigma_{\hat{M}_{500}}^{\mathrm{NFW+2h}}$ for the fitted truncated NFW+2h convergence model and which we use as a metric of constraining power.

We find that at a given point in true cluster parameter ($M_{500}$--$z$) space, $\sigma_{\hat{M}_{500}}^{\mathrm{NFW}}$ and $\sigma_{\hat{M}_{500}}^{\mathrm{NFW+2h}}$ are within a few percent of each other. The exact value of their ratio as a function of $M_{500}$ is quite sensitive to the details of the polynomial fit to $\beta_{\mathrm{c}}$ and $\sigma_{\mathrm{c}}$ (specifically, the polynomial degree and the weighting of each bin), especially at the high-mass end. However, we consistently find some trends. For $z=0.23$, $\sigma_{\hat{M}_{500}}^{\mathrm{NFW+2h}}$ is about 1--2\,\% larger than $\sigma_{\hat{M}_{500}}^{\mathrm{NFW}}$ at low masses, and about $0.5$--1\,\% smaller at intermediate and high masses. That is, the original truncated NFW model slightly outperforms the fitted truncated NFW+2h model at low masses, but becomes less optimal at higher masses. We argue that this is due to the intrinsic scatter $\sigma_{\mathrm{c}}$ being larger at low masses for the fitted truncated NFW+2h model (see Figs~\ref{fig:edgeworth_full} and \ref{fig:twohalo}).
%\AC{The following could be removed if now discussed earlier: which we attribute to the fact that it is more spatially extended, and therefore more sensitive to variations around the mean convergence due to correlated LSS.} 
At higher masses, however, the scatters take similar values, and therefore the small difference in mean signal-to-noise (see Fig.~\ref{fig:ratio_mean}), which favours the fitted truncated NFW+2h model, becomes the main source of difference in constraining power. For $z=0.52$, the results are more sensitive to the fitting details, since there are fewer clusters, especially at the high-mass end, where we cannot determine any trend conclusively. At low masses, however, we find $\sigma_{\hat{M}_{500}}^{\mathrm{NFW+2h}}$ to be about 5--10\,\% larger than $\sigma_{\hat{M}_{500}}^{\mathrm{NFW}}$, and about 1--3\,\% at intermediate masses. As at $z=0.23$, we attribute this to the (now much larger) intrinsic scatter difference between the two models (see Figs~\ref{fig:edgeworth_full} and \ref{fig:twohalo}).

In summary, we find that, at low masses, the fitted truncated NFW+2h convergence model, despite yielding higher signal-to-noise measurements, performs slightly worse than the truncated NFW convergence model due to it having a larger associated intrinsic scatter. For $z=0.23$, the situation is reversed at intermediate and high masses, where the intrinsic scatter is similar. These differences, however, are small, at the few percent level.

\subsubsection{Priors in \citet{Zubeldia2019}}

%Justification of priors, and justification of neglect of higher-order corrections.

As already noted in Section \ref{subsubsec:amethod}, the approach to understanding the observations considered in this section underlies the hierarchical model of \citet{Zubeldia2019}, where $1-b_{\mathrm{CMBlens}}$, $\sigma_{\mathrm{CMBlens}}$, and the variable $p_{\text{t}}$ correspond to $\beta_{\mathrm{c}}$, $\sigma_{\mathrm{c}}$, and $p_{\mathrm{c}}$ here, respectively. We note, however, that  $1-b_{\mathrm{CMBlens}}$ in \citet{Zubeldia2019} is also intended to account for the few-percent negative bias due to cluster miscentering. In that work, the intrinsic scatter is assumed to be log-normal, which, as shown in Section \ref{sufficiency}, is a sufficiently good approximation, and priors are imposed on $1-b_{\mathrm{CMBlens}}$ and $\sigma_{\mathrm{c}}$. These are Gaussian priors centred at $1-b_{\mathrm{CMBlens}} = 0.93$ and $\sigma_{\mathrm{CMBlens}}=0.2$, both with standard deviation of $0.05$.

The cluster sample used in \citet{Zubeldia2019} is the \textit{Planck} MMF3 cosmology sample, which consists of 439 SZ-detected clusters (see \citealt{Planck2016xxvii}). In it, most clusters are at low redshift (see Fig.~1 of \citealt{Planck2016xxvii}), with 246 out of 433 clusters with measured redshift being at $z<0.23$ ($z=0.23$ being the redshift of our low-redshift snapshot), and 311 being at $z<0.3$. Only 22 clusters are at $z>0.52$ (i.e., above the redshift of our high-redshift snapshot). Also, in Fig.~1 of \citet{Planck2016xxvii} it can be seen that most cluster masses take values lying around the centre of our considered mass range at lower redshifts, and values towards our high-mass end at higher redshifts. The latter is particularly true for clusters with redshifts close to our high-redshift snapshot. Therefore, most of the clusters in the sample lie within the region of mass--redshift parameter space where the priors imposed on $1-b_{\mathrm{CMBlens}}$ and $\sigma_{\mathrm{CMBlens}}$ in \citet{Zubeldia2019} are consistent with our measurements of $\beta_{\mathrm{c}}$ (plus a few percent decrement due to miscentering; see \citealt{Zubeldia2019}) and $\sigma_{\mathrm{c}}$, respectively. 

%We also argue that the deviations from log-normality that we find in this work and that are not taken into account in \citet{Zubeldia2019} can be safely neglected. Indeed, as it can be seen in Fig \ref{fig:histograms}, the log-normal model provides a rather good fit to our mock observations, which only include a small amount of scatter due to uncorrelated LSS and which do not include the scatter due to reconstruction noise. Since this additional scatter, which has been argued above to be approximately Gaussian, is much larger than the intrinsic scatter, the mismatch between the log-normal model and the fully noisy measurements ought to be even smaller. We note, nevertheless, that future experiments with lower reconstruction noise may have to take any potential deviations from log-normality in their mass observable into account.

\subsection{Extrapolation approach}\label{subsec:b}

\subsubsection{Statistics as a function of integration length}\label{subsubsec}

In the extrapolation approach, only the reconstruction noise power spectrum is present in the matched-filter inverse-variance weighting (see Eq.~\ref{filtval}). As in Section \ref{subsec:b}, let us first think of a hypothetical set of simulated observations with reconstruction noise and with LSS present from $z=0$ back to last-scattering. Since reconstruction noise is approximately Gaussian, as a consequence of our choice of matched-filter weighting $P(p_{\mathrm{obs}}|p,M_{500},z;\chi_{\star})$ is approximately a unit-variance Gaussian centred on $p$. We remind that $p = p_{\mathrm{c}}+\Delta p_{\mathrm{u}}$, and that $\chi_{\star}$ denotes that integration is performed from $z=0$ back to last-scattering. In order to fully determine the statistics of $p_{\mathrm{obs}}$ at given true cluster parameters, $M_{500}$ and $z$, what remains to be characterised is $P(p|M_{500},z;\chi_{\star})$. In the context of this approach we refer to the variability due to reconstruction noise, $P(p_{\mathrm{obs}}|p,M_{500},z;\chi_{\star})$, as \textit{observational} scatter, and to the remaining variability, due to the cluster itself, correlated and uncorrelated LSS, $P(p|M_{500},z;\chi_{\star})$, as \textit{intrinsic} scatter. We note that this way of understanding the scatter is different from that made in Section \ref{subsec:a}, in which the variation due to uncorrelated LSS is understood as being part of the observational scatter. 

The aim of the rest of this section is to determine $P(p|M_{500},z;\chi_{\star})$, which we do with our mock observations, namely with our cluster and random data sets $B$. We first study the statistics of $p$ as a function of $M_{500}$, $z$, and integration length, $\chi_l$. We can do this with the mock observations of cluster data set $B$, which can be thought of as samples from $P(p,M_{500}|z;\chi_l)$, where $\chi_l$ can be any of the seven integration lengths that we consider; $P(p|M_{500},z;\chi_l)$ can be obtained as slices in $M_{500}$ of $P(p,M_{500}|z;\chi_l)$. In particular, for each of our two snapshot redshifts and each of our seven values of $\chi_l$, we divide our data points into 90 subsets, binning $M_{500}$ in 90 equally-spaced bins between $2 \times 10^{14} M_{\odot}$ and $10^{15} M_{\odot}$. As in Section \ref{subsec:a}, we think of the value of $p$ of each data point falling within a given bin as a sample of $P(p|M_{500},z;\chi_l)$, $M_{500}$ being the mass at the bin centre. 

In order to characterise the distribution $P(p|M_{500},z;\chi_l)$, we first compute its empirical mean, which we denote with $\left\langle p \right\rangle$, with the values of $p$ within each mass bin. We also compute the empirical mean, standard deviation, skewness, and kurtosis of $P(\ln p|M_{500},z;\chi_l)$, which we denote with $\mu$, $\sigma$, $\lambda_3$, and $\lambda_4$, respectively, as a function of $M_{500}$ and $\chi_l$ for our two snapshots. Analogously to what is done in Section \ref{subsec:a}, $\lambda_{3}$ and $\lambda_{4}$ are defined in terms of the cumulants of $P(\ln p|M_{500},z;\chi_l)$, $k_n$, as $\lambda_{3} = k_3/\sigma^3$ and $\lambda_{4} = k_4/\sigma^4$, respectively. Also as in Section \ref{subsec:a}, instead of presenting our results in terms of $\mu$, we introduce a bias parameter, $\beta$, which is defined in an analogous way to $\beta_{\mathrm{c}}$ in Section \ref{subsec:a}, i.e., by imposing that $\ln \bar{p}(\beta M_{500},z)=\mu$, where $\bar{p}(\beta M_{500},z)$ is given by Eq. (\ref{mfsnr}), noting that $\sigma_{M_{500}}$ is now the noise of the matched filter with only reconstruction noise in the inverse-variance weighting.

\begin{figure}
\centering
\includegraphics[width=0.5\textwidth]{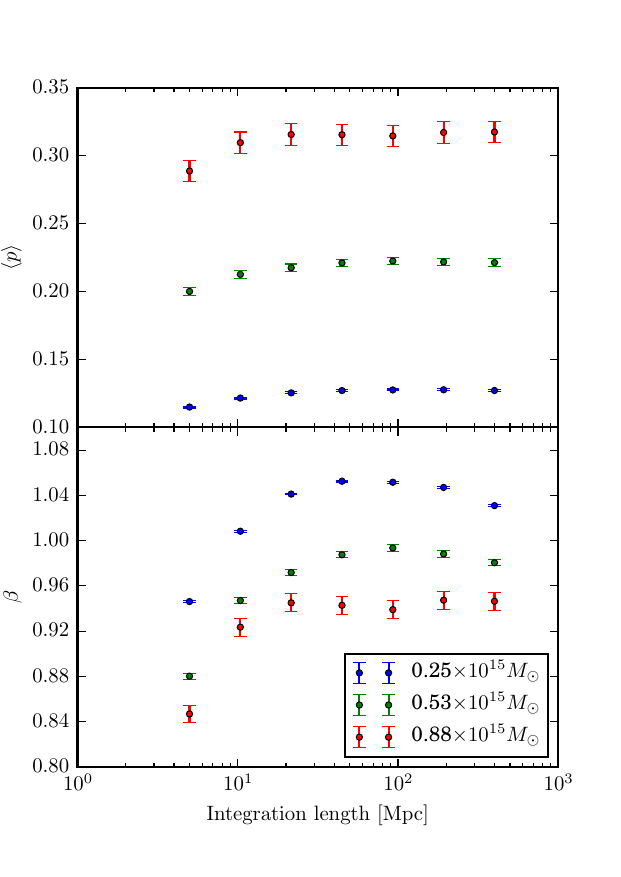}
\caption{Expected value of $p$, $\left\langle p \right\rangle$, (upper panel) and bias, $\beta$, (lower panel) as a function of integration length, $\chi_l$, (in comoving Mpc) for three representative $M_{500}$ bins, as obtained from the cluster data set $B$ (see Section \ref{subsec:b}). The error bars are obtained with bootstrapping.}
\label{fig:results_lz_fixed_mass}
\end{figure}

\begin{figure*}
\centering
\includegraphics[width=0.9\textwidth]{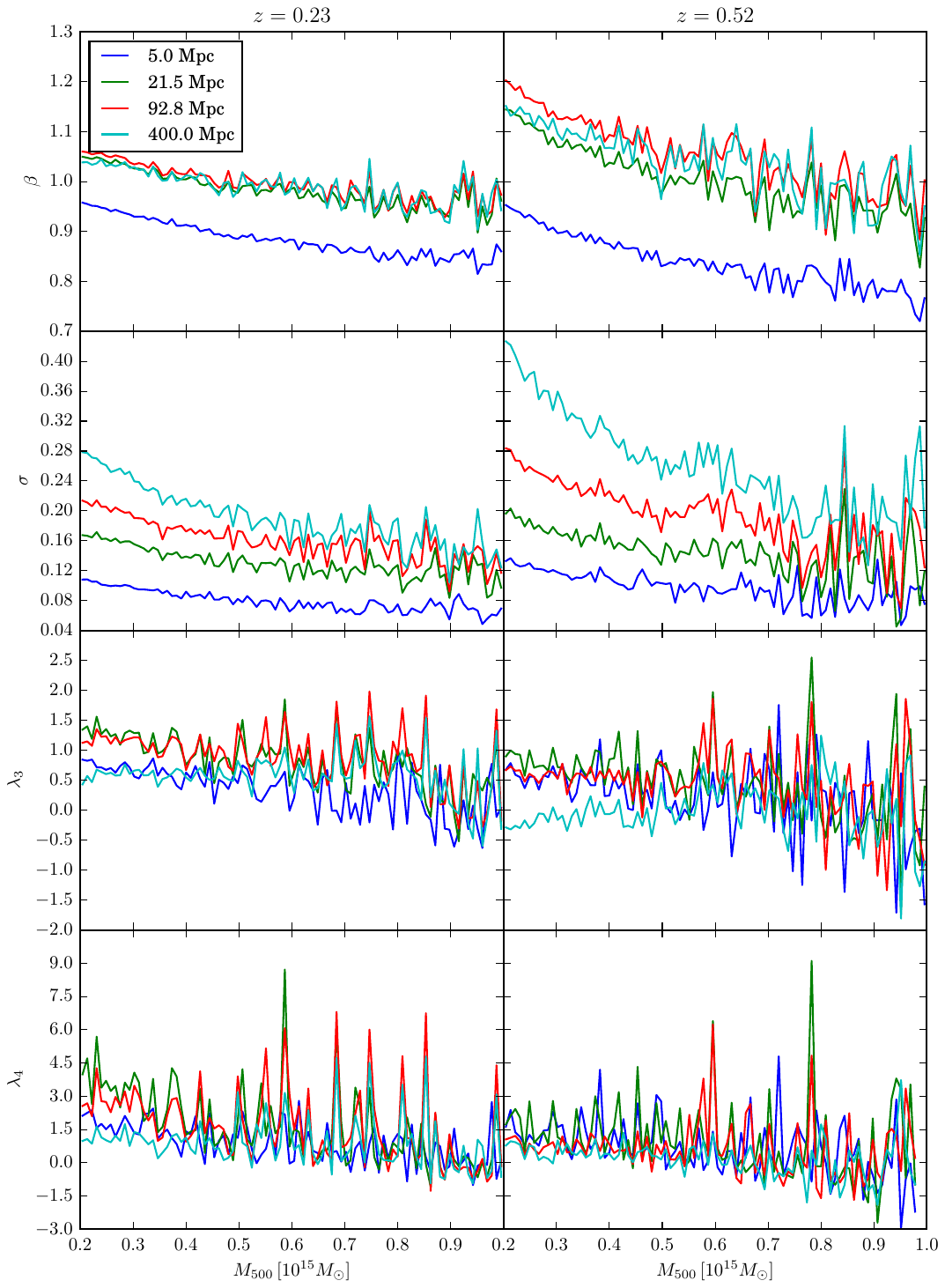}
\caption{Measured values of $\beta$ (bias), $\sigma$ (scatter), $\lambda_{3}$ (skewness), and $\lambda_{4}$ (kurtosis) as a function of $M_{500}$ and integration length, $\chi_l$, for $z=0.23$ (left) and $z=0.52$ (right), as obtained from cluster data set $B$ (see Section \ref{subsec:b}).}
\label{fig:results_lz}
\end{figure*}

Figure \ref{fig:results_lz_fixed_mass} shows $\left\langle p \right\rangle$ and $\beta$ as a function of integration length for three selected mass bins for the $z=0.23$ snapshot. We obtain the error bars with bootstrapping, resampling from our cluster data points $(M_{500},p)$, with a $(M_{500},p)$ pair taken as a single data point so that the correlation structure is not lost. As expected, the errors are larger for higher masses, as there are fewer clusters per bin. In addition, our measurements of $\beta$, $\sigma$, $\lambda_3$, and $\lambda_4$ are shown in Fig.~\ref{fig:results_lz} as a function of $M_{500}$ and integration length for our two redshifts. For the sake of clarity, only four integration lengths have been included.

Since $p$ is linear in the lensing convergence (see Eq. \ref{pdef}), $\left\langle p \right\rangle$ is linear in the mean lensing convergence within a given integration length at given true cluster parameters and is insensitive to other moments of the convergence. This is illustrated in the upper panel of Fig.~\ref{fig:results_lz_fixed_mass}: $\left\langle p \right\rangle$ responds to the change in the mean convergence as the integration length is increased starting from its lowest value, increasing as more correlated LSS is added, but this growth plateaus at large integration lengths, where effectively only zero-mean uncorrelated LSS is being added.

The bias parameter $\beta$, however, defined in terms of $\ln p$, is not linear in the convergence. In general it feels both the effect of the mean convergence at a given integration length and of the variance, and higher moments, around the mean convergence. In Section \ref{subsec:resultsdisc} we saw that the analogous parameter defined in the context of the deconvolution approach, $\beta_{\mathrm{c}}$, is effectively only sensitive to the mean convergence, due to the fact that the intrinsic scatter is small; that is, this analogous parameter essentially corrects the mismatch between the true mean convergence and the model used in the matched filter. Following an argument similar to that in Section \ref{subsec:resultsdisc}, and assuming log-normality  (as shown in 
Eq.~\ref{edgeworthmean}, the effect of the skewness and the kurtosis is higher order in the scatter and therefore negligible),
\begin{equation}\label{lognormalmean}
\ln \bar{p}(\beta M_{500},z) = \ln \left\langle p \right\rangle - \frac{\sigma^2}{2} \, .
\end{equation}{}
At low integration lengths, $\sigma$ is small, as can be seen in Fig.~\ref{fig:results_lz}, and, as in the analogous case in Section \ref{subsec:resultsdisc}, negligible. The bias, $\beta$, is therefore just sensitive to the mean convergence profile through $\ln \left\langle p \right\rangle$, which at low integration lengths changes with integration length, as there is still correlated LSS that is being added. In particular, we can see in Fig.~\ref{fig:results_lz_fixed_mass} that $\beta$ initially increases with integration length. As the integration length increases, however, the mean convergence changes progressively less, which can be seen in the plateauing of $\left\langle p \right\rangle$ in Fig.~\ref{fig:results_lz_fixed_mass}, and more uncorrelated LSS is added, with $\sigma$ increasing as a result (see Fig.~\ref{fig:results_lz}). Equation~(\ref{lognormalmean}) shows that at fixed $\left\langle p \right\rangle$, if $\sigma$ increases $\bar{p}$ decreases and, as a consequence, so does $\beta$. This fall with integration length is observed at high integration lengths in Figs~\ref{fig:results_lz_fixed_mass} and \ref{fig:results_lz}, being more pronounced at lower masses.

In Fig.~\ref{fig:results_lz_fixed_mass} it can also be seen that $\beta$ and $\sigma$ decrease with $M_{500}$ for both redshifts, analogously to $\beta_{\mathrm{c}}$ and $\sigma_{\mathrm{c}}$ in Section \ref{subsec:a}. In addition, the skewness, $\lambda_3$, and kurtosis, $\lambda_4$, exhibit a similar qualitative behaviour as a function of integration length across mass and for both redshifts: they both take lower values for $\chi_l=5$\,Mpc, the shortest integration length considered, grow with $\chi_l$ to reach some maximum value, and then decrease beyond this as effectively only uncorrelated LSS is added. That is, above some value of $\chi_l$, as integration length increases the intrinsic scatter, due to the cluster itself, correlated and uncorrelated LSS, becomes increasingly log-normal. As in Section \ref{subsec:a}, the skewness and the kurtosis are generally lower for $z=0.52$ than for $z=0.23$.

\subsubsection{Extrapolation}

The fact that for large values of $\chi_l$, at which effectively only uncorrelated LSS is added on increasing $\chi_l$ further, $P(p|M_{500},z;\chi_l)$ becomes increasingly log-normal motivates the following approach to obtain $P(p|M_{500},z;\chi_{\star})$, the ultimate goal of this section. 

For a given integration length $\chi_l$, $p = p_{\textrm{c}} + \Delta p_{\textrm{u}}$. Since  $p_{\textrm{c}}$ and $\Delta p_{\textrm{u}}$ are independent random variables, the cumulants of $p$ are equal to the cumulants of $p_{\textrm{c}}$ plus the cumulants of $\Delta p_{\textrm{u}}$. In particular, the mean of $p$, $\left\langle p \right \rangle$, is equal to the mean of $p_{\textrm{c}}$, since $\Delta p_{\textrm{u}}$ has zero mean, and $\sigma_p^2 = \sigma_{p_{\mathrm{c}}}^2 + \sigma_{\Delta p_{\textrm{u}}}^2$, where $\sigma_p^2$, $\sigma_{p_{\mathrm{c}}}^2$, and $\sigma_{\Delta p_{\textrm{u}}}^2$ are the variances of $p$, $p_{\textrm{c}}$, and $\Delta p_{\textrm{u}}$, respectively.

For $\chi_l = 400$\,Mpc we measure $\left\langle p \right \rangle $ and $\sigma_p^2$ from our cluster data set $B$ and $\sigma_{\Delta p_{\textrm{u}}}^2$ from our random data set $B$ across the 90 $M_{500}$ bins. We then obtain $\sigma_{p_{\mathrm{c}}}^2$ by subtraction. Since this integration length effectively contains all LSS correlated with the cluster, our measured values of $\left\langle p \right \rangle$ and $\sigma_{p_{\mathrm{c}}}^2$ are the same as the values of such cumulants for $\chi_l = \chi_{\star}$, which we denote with $\left\langle p \right \rangle (\chi_{\star})$ and $\sigma_{p_{\mathrm{c}}}^2 (\chi_{\star})$, respectively. Thus, we just need to determine $\sigma_{\Delta p_{\textrm{u}}}^2(\chi_{\star})$, the variance associated with LSS from $z=0$ to last-scattering.

In general, $\sigma_{\Delta p_{\textrm{u}}}^2(\chi_l)$ can be written as
\begin{equation}\label{predictedvariance}
\sigma_{\Delta p_{\textrm{u}}}^2(\chi_l) = \left[\int \frac{d^2 \bmath{L}}{2 \upi} \frac{|\kappa_{\textnormal{t}} (\bmath{L})|^2}{N_{\kappa}(\bmath{L})} \right]^{-1} \int \frac{d^2 \bmath{L}}{2 \upi} \frac{|\kappa_{\textnormal{t}} (\bmath{L})|^2}{|N_{\kappa}(\bmath{L})|^2} C_L^{\kappa \kappa}(\chi_l) \, ,
\end{equation}
where $C_L^{\kappa \kappa}(\chi_l)$ is the power spectrum of the lensing convergence within the comoving distance interval specified by $\chi_l$, and the other variables are the same as in Eq. (\ref{pdef}). We compute $C_L^{\kappa \kappa}(\chi_l)$ with CAMB\footnote{\texttt{www.camb.info}} using the nonlinear matter power spectrum and the Limber approximation. We find that for $\chi_l = 400$\,Mpc and for $\chi_l = 192.7$\,Mpc, $\sigma_{\Delta p_{\textrm{u}}}^2(\chi_l)$ computed this way is within a few percent of the value we measure across our $M_{500}$ bins and our two redshifts. We also find that, for a given redshift and mass bin, the small fractional disagreement for $\chi_l = 400$\,Mpc is very similar to that for $\chi_l = 192.7$\,Mpc. This motivates us to compute $\sigma_{\Delta p_{\textrm{u}}}^2(\chi_{\star})$ using Eq.~(\ref{predictedvariance}) with a simple rescaling in each mass bin and at each redshift by the appropriate factor obtained empirically for $\chi_l = 400$\,Mpc. We add this rescaled value of $\sigma_{\Delta p_{\textrm{u}}}^2(\chi_{\star})$ to $\sigma_{p_{\mathrm{c}}}^2(\chi_{\star})$ to obtain finally $\sigma_{p}^2(\chi_{\star})$

We then assume that $P(p|M_{500},z;\chi_{\star})$ is log-normal. Denoting the mean and standard deviation of $P(\ln p|M_{500},z;\chi_{\star})$ with $\mu_{\mathrm{e}}$ and $\sigma_{\mathrm{e}}$, respectively, we compute them by imposing that
the mean and variance of $P(p|M_{500},z;\chi_{\star})$ have the values $\left\langle p \right \rangle (\chi_{\star})$ and $\sigma_{p}^2(\chi_{\star})$ that we determine as described above. Following Sections \ref{subsec:a} and \ref{subsubsec}, instead of presenting our results in terms of $\mu_{\mathrm{e}}$ we introduce a bias parameter, $\beta_{\mathrm{e}}$, defined such that $\ln \bar{p}(\beta_{\mathrm{e}}M_{500},z) = \mu_{\mathrm{e}} $, with $\bar{p}$ given by Eq. (\ref{mfsnr}) (where $\sigma_{M_{500}}$ is now computed with only the reconstruction noise power spectrum in the inverse-variance matched-filter weighting).

\begin{figure}
\centering
\includegraphics[width=0.45\textwidth]{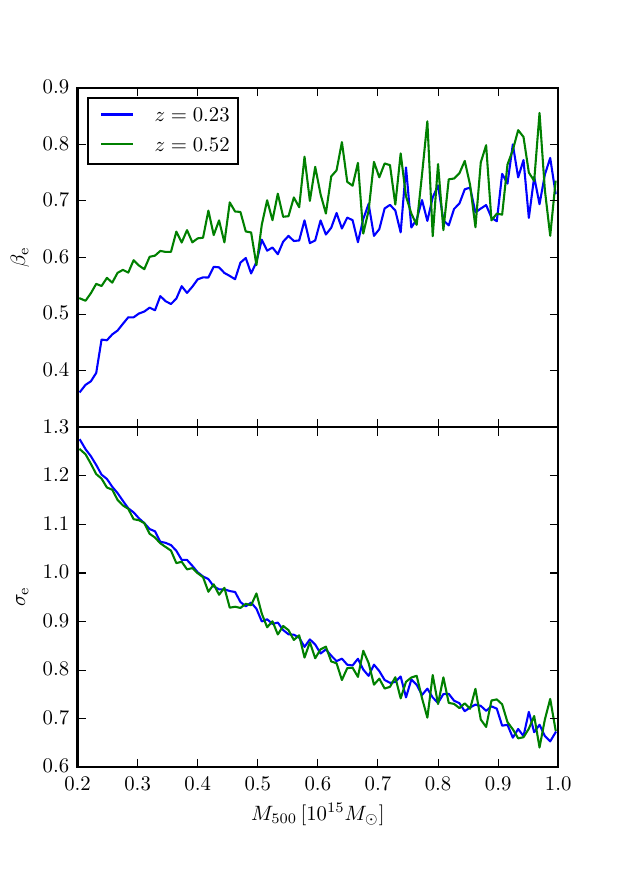}
\caption{Measured values of the lensing mass bias, $\beta_{\mathrm{e}}$ (upper panel), and scatter, $\sigma_{\mathrm{e}}$ (lower panel), extrapolated to include uncorrelated LSS from $z=0$ to last-scattering for our \textit{Planck}-like reference experiment as a function of $M_{500}$ and for $z=0.23$ (blue curves) and $z=0.52$ (green curves).}
\label{fig:results_extrapolated}
\end{figure}

%Description of results

The bias, $\beta_{\mathrm{e}}$, and scatter, $\sigma_{\mathrm{e}}$, obtained this way are shown in Fig.~\ref{fig:results_extrapolated}. As expected, the addition of a large amount of uncorrelated LSS significantly decreases
$\beta_{\mathrm{e}}$ compared to our previous measurements of $\beta$ for smaller integration lengths (see Fig.~\ref{fig:results_lz} and Eq.~\ref{lognormalmean}). This is true across our $M_{500}$ range and for both redshifts, the effect being more significant at lower masses. Similarly, the scatter has increased substantially with respect to the values observed for smaller integration lengths (see also Fig.~\ref{fig:results_lz}), being dominated by uncorrelated LSS.\footnote{We note that $\sigma_{\mathrm{e}}$ takes similar values for the two snapshot redshifts across the mass range considered. This seems to be a coincidence specific to our choice of matched filter convergence template. Indeed, to investigate this, we have obtained $\sigma_{\mathrm{e}}$ as a function of $M_{500}$ for other choices of convergence model (e.g., our fitted truncated NFW+2h model), and found that, in general, there is a dependency on redshift.}

%\AC{Inigo, if I've got things correct,
%
%\begin{equation*}
% e^{\sigma_{\text{e}}^2} -1 = \frac{\sigma_{p}^2(\chi_{\star})}{\langle p\rangle^2(\chi_\star)} \, . 
%\end{equation*}
%
%We expect $\langle p\rangle(\chi_\star)$ to be approximately proportional to mass, while the variance to be almost independent of mass since dominated by uncorrelated LSS and the change in filter with mass has little impact on $\sigma^2_{\Delta p_{\text{u}}}(\chi_\star)$ (expect it will decrease slightly with increasing mass due to the larger filter scale, but the $N^{-1}_{\kappa}(L)$ suppresses this change). It follows that $e^{\sigma_{\text{e}}^2}}-1$ should go like $1/M^2_{500}$. Furthermore, since
%
%\begin{equation*}
%\langle p \rangle = \bar{p}(\beta_{\text{e}}M_{500}) e^{\sigma_{\text{e}}^2/2} \, ,
%\end{equation*}
%
%and both \langle p \rangle$ and $\bar{p}(\beta_{\text{e}}M_{500})$ are approximately proportional to mass, we expect $\beta_{\text{e}}$ to be proportional to $e^{-\sigma^2_{\text{e}}/2$ and so to increase with mass. What I am a bit surprised by is the approximate independence of $\sigma_{\text{e}}$ on $z$ in Fig.~\ref{fig:results_extrapolated}. I would have expected $\langle p \rangle(\chi_\star)$ to decrease with redshift at given $M_{500}$ but $\sigma_{p}(\chi_\star)$ to be almost independent, in which case $\sigma_{\text{e}}^2$ would be expected to increase with redshift, which isn't the case. Any thoughts?}

%

We remark that this extrapolation approach relies on the assumption of log-normality of $P(p|M_{500},z;\chi_{\star})$, an assumption that, although motivated by our mock observations (see, e.g., Fig.~\ref{fig:results_lz}), we cannot check empirically. A more rigorous approach to study the full line-of-sight intrinsic scatter would involve considering full lightcone simulations of the lensing convergence from $z=0$ back to last-scattering (see, e.g., \citealt{Giocoli2016,Takahashi2017}), something that is beyond the scope of this paper.

%Pros: Better model of total LSS (regarding future experiments), correlation with SZ. %Cons: extrapolation

In summary, this extrapolation approach treats the total scatter due to the cluster and to correlated and uncorrelated LSS as log-normal, and reconstruction noise as Gaussian. In contrast, the deconvolution approach (Section \ref{subsec:a}) treats the scatter due to the cluster and to correlated LSS as log-normal (or close to log-normal), and the scatter due to uncorrelated LSS and reconstruction noise as Gaussian. For an experiment like \textit{Planck}, both approaches are essentially equivalent. Indeed, as shown in Section \ref{subsec:a}, the scatter due to the cluster and to correlated LSS is approximately log-normal, at least for mid to high-mass clusters, and the scatter due to uncorrelated LSS and reconstruction noise is roughly Gaussian, since reconstruction noise is almost Gaussian and dominates over uncorrelated LSS (which itself is approximately Gaussian on the scales relevant for cluster mass estimation with \textit{Planck}). In addition, as argued in this section, the total scatter due to the cluster and to correlated and uncorrelated LSS is approximately log-normal for a \textit{Planck}-like experiment.

This equivalence may no longer be valid for future experiments, which will have lower reconstruction noise, i.e., higher signal-to-noise per cluster, and probe the convergence on smaller angular scales. If reconstruction noise becomes comparable with the scatter due to uncorrelated LSS, approximating the summed scatter as Gaussian may no longer be accurate. The extrapolation approach, therefore, may be preferable in such cases. We note that the extrapolation approach does not include the LSS convergence power spectrum in the inverse-variance weighting of the matched filter, and, as a consequence, does not optimise for it, as opposed to the deconvolution approach. There is therefore some signal-to-noise loss in the extrapolation approach with respect to the deconvolution approach. For \textit{Planck}, which is reconstruction-noise dominated, this loss is small, but it may be more significant in lower reconstruction noise scenarios.

\section{Different observational specifications}\label{sec:different}

\begin{figure}
\centering
\includegraphics[width=0.45\textwidth]{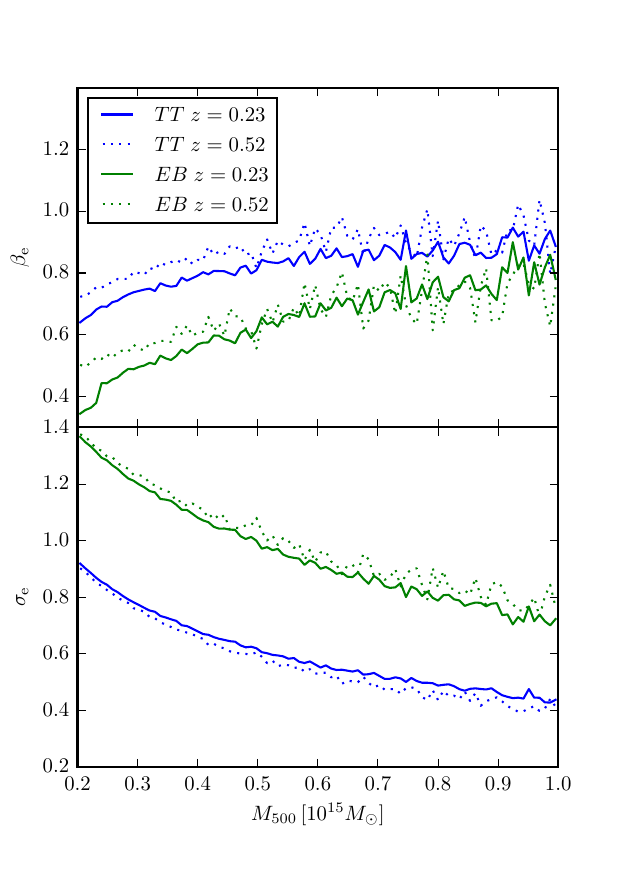}
\caption{Measured values of the lensing mass bias, $\beta_{\mathrm{e}}$ (upper panel), and scatter, $\sigma_{\mathrm{e}}$ (lower panel), extrapolated to include uncorrelated LSS from $z=0$ to last-scattering, for an AdvACT-like experiment with the $TT$ quadratic estimator (in blue) for $z=0.23$ (solid curves) and $z=0.52$ (dotted curves). Analogous results for the $EB$ quadratic estimator are shown in green.}
\label{fig:results_extrapolated_act}
\end{figure}

We briefly consider how the results obtained with our extrapolation approach vary if we assume a different experimental setup. In particular, we consider an idealised experiment analogous to the \textit{Planck}-like experiment considered throughout this paper but with a FWHM = 1.4\,arcmin and with temperature noise levels of 7\,$\mu$K\,arcmin; this is intended to resemble AdvACT. We consider two different lensing quadratic estimators, the $TT$ quadratic estimator, which is the one used throughout this paper for our \textit{Planck}-like experiment, and the $EB$ quadratic estimator. For both quadratic estimators we use a maximum multipole of $l=2000$ in the gradient leg and of $l=3000$ in the field leg. 

The $EB$ quadratic estimator differs from the $TT$ quadratic estimator in that it takes as input one map of the (lensed) $E$-mode and one map of the (lensed) $B$-mode CMB polarisation in order to estimate the lensing convergence, instead of two copies of the (lensed) temperature anisotropies. (See~\citealt{Hu2002} for the detailed construction of the $EB$ estimator.) The $E$-modes and $B$-modes are a useful basis-independent description of the (linear) polarisation of the CMB~\citep{Seljak1997,KKS1997}. Lensing reconstruction with polarisation is significantly noisier than reconstruction with temperature for an experiment like \textit{Planck} \citep{Planck2018VIII}. Indeed, for this experiment, for a cluster with $M_{500} = 5 \times 10^{14} M_{\odot}$ at redshift $z=0.23$ and with the convergence profile of our matched-filter template (Eq.~\ref{template}), we find, using Eq. (\ref{mfsnr}) with $\beta_{\mathrm{c}}=1$, a mean signal-to-noise of $\left\langle p \right\rangle = 0.21$ if the $TT$ quadratic estimator is used. If the $EB$ quadratic estimator is used instead, we find $\left\langle p \right\rangle = 0.03$.
%That is the reason why in this paper we have not considered the $EB$ quadratic estimator for the \textit{Planck}-like reference experiment.
The situation is, however, different for our AdvACT-like experiment, for which, for the same cluster, we find $\left\langle p \right\rangle = 0.42$ if the $TT$ estimator is used, and $\left\langle p \right\rangle = 0.41$ if the $EB$ estimator is used. The $EB$ estimator therefore provides a signal-to-noise similar to that of the $TT$ estimator for this experimental set-up. It also has the additional interest that it is less sensitive to extragalactic foregrounds than the $TT$ estimator. In the cluster context, the $TT$ estimator in particular suffers from contamination due to the thermal and kinetic SZ effects from the cluster itself, the latter being difficult to remove given that it has the same frequency dependence as the CMB anisotropies, while the polarisation signal from such effects is a negligible contaminant~\citep{Raghunathan2017}. We note that the other three quadratic estimators, the $TE$, the $TB$, and the $EE$ estimators, yield, in this scenario, smaller mean signal-to-noise ratios, $\left\langle p \right\rangle = 0.22$, $\left\langle p \right\rangle = 0.14$, and $\left\langle p \right\rangle = 0.24$, respectively. We therefore do not consider them here.

The extrapolated bias, $\beta_{\mathrm{e}}$, and scatter, $\sigma_{\mathrm{e}}$, that we obtain using the $TT$ and $EB$ quadratic estimators for our AdvACT-like set-up are shown in Fig.~\ref{fig:results_extrapolated_act}. It can be observed that the values of  $\beta_{\mathrm{e}}$ and $\sigma_{\mathrm{e}}$ as a function of $M_{500}$ and $z$ depend on the choice of quadratic estimator and, also, on the experiment specifications (compare with Fig.~\ref{fig:results_extrapolated}, where analogous results are shown for the \textit{Planck}-like experiment). This implies that, although the calibration approaches presented in this paper may be applied to other lensing estimation techniques and experimental set-ups, their numerical results require case-by-case consideration.

\section{Conclusion}\label{sec:conclusions}

%Summary of why this is useful, why CMB lensing is useful, what we've done for Planck, two different approaches, relation to Zubeldia & Challinor (2019)

We have studied the statistics of a CMB lensing galaxy cluster mass observable, $p$, for a \textit{Planck}-like experiment with mock observations obtained from an $N$-body simulation, characterising the biased mean signal and the scatter, and deviations from log-normality, due to the variation associated with the cluster and with correlated and uncorrelated LSS. This characterisation is essential for a cosmological analysis that may make use of this mass observable (e.g., \citealt{Zubeldia2019}) to deliver unbiased results.

We have followed two alternative routes in order to quantify the statistics of this mass observable. First, in our deconvolution approach (Section \ref{subsec:a}), we have treated the variation due to uncorrelated LSS as noise in the matched-filtering process, where it adds to the reconstruction noise, and then characterised the mean signal and the variation due to the cluster itself and to correlated LSS (what we call intrinsic scatter) with our mock observations. We find this intrinsic scatter to be roughly log-normal, although significant skewness and kurtosis are detected, as in similar studies of galaxy weak lensing cluster mass measurements (e.g., \citealt{Becker2011}, where significant positive skewness is found). We find, however, these deviations from log-normality to have a negligible impact on mass calibration for our \textit{Planck}-like experiment. This approximate log-normality and our measured values of the bias, $\beta_{\mathrm{c}}$, and the scatter, $\sigma_{\mathrm{c}}$, serve as justification for the priors on the analogous parameters used in \citet{Zubeldia2019}. Second, in our extrapolation approach (Section \ref{subsec:b}) we have considered the variation due to the cluster itself and to both correlated and uncorrelated LSS as intrinsic scatter. We find with our mock observations that this scatter becomes increasingly log-normal as we integrate along longer paths, which motivates us to extrapolate our results in order to incorporate the scatter due to uncorrelated LSS from $z=0$ back to CMB last-scattering assuming log-normality.

%Other surveys: statistics vary, each survey has to calibrate theirs.

We have also considered, for illustration, how our extrapolation results change if a different experimental set-up is assumed (an AdvACT-like experiment; see Section \ref{sec:different}). While the qualitative trends with mass and redshift are similar, numerical results for $\beta_{\text{e}}$ and $\sigma_{\text{e}}$ differ significantly from those obtained for the \textit{Planck}-like experiment. This implies that ongoing and future experiments that may want to use our CMB lensing mass observable for, e.g., mass calibration in a cluster counts analysis, will have to quantify its statistics for the particular case of their experiment specifications. Our numerical results, for both a \textit{Planck}-like and an AdvACT-like experiment, are not transferable to other experiments. Different cluster mass observables (e.g., the observable proposed in \citealt{Raghunathan2017} or in \citealt{Horowitz2017}) are also expected to have different statistics and would also require custom calibration if they were to be used in a cosmological analysis.

%Future things to investigate: Correlations with other proxies. Impact of baryons, systematics very difficult to quantify. Impact of deviations from assumed log-normality, and in particular skewness, on cosmology. Impact of getting scatter/bias wrong on cosmology. Full line of sight simulations

As demonstrated in this work, simulations provide a useful means to quantify the statistics of CMB lensing cluster mass observables. Future work, however, will be needed in order to improve upon and extend our results on several fronts, for this CMB lensing observable and for others. First, the impact of baryonic effects on our results is difficult to quantify. Simulations that incorporate baryonic effects may be useful in this respect, although currently the number of massive galaxy clusters produced in such simulations is not large enough in order to compete statistically with the results presented in this paper. Indeed, the number of galaxy clusters from state-of-the art simulations with baryons is only around 10--100 \citep{Barnes2016,Planelles2017,Truong2017,Henden2019}, significantly lower than our number of clusters, around $10^4$. Furthermore, the possible dependence of our results on cosmological parameters has yet to be determined. The impact of deviations from log-normality on the mass calibration of a cluster sample and, in turn, on the cosmological constraints drawn from such sample also remains to be investigated (potentially along the lines of, e.g., \citealt{Shaw2010}). We have argued that this is a negligible effect for the \textit{Planck} galaxy clusters, but it may not be negligible for future CMB experiments, which will deliver higher signal-to-noise CMB lensing cluster mass measurements. The correlations of CMB lensing mass observables with other cluster observables (e.g., the SZ and X-ray signals) also need to be quantified; simulations with baryons may also be useful in this respect. Future CMB experiments such as CMB-S4~\citep{Abazajian2016} will enable SZ counts analyses in which the SZ--mass scaling relations will be able to be calibrated completely with CMB lensing masses alone to sub-percent accuracy \citep{2017Louis}. If their full statistical power is to be realised without biases, assessing the impact of these potential issues and accurately determining biases, scatter and intrinsic correlations will be an essential step for these studies.

\section*{Acknowledgements}

The authors would like to thank Debora Sijacki, William Handley, and, especially, Ewald Puchwein for useful discussions.
IZ is supported by the Isaac Newton Studentship from the University of Cambridge. AC acknowledges support from the UK Science and Technology Facilities Council (grant numbers ST/N000927/1 and ST/S000623/1).
Part of the computational work of this paper was done using the Cambridge Service for Data Driven Discovery (CSD3) operated by the University of Cambridge Research Computing Service (\url{http://www.csd3.cam.ac.uk/}), provided by Dell EMC and Intel using Tier-2 funding from the Engineering and Physical Sciences Research Council, and DiRAC funding from the Science and Technology Facilities Council (\url{www.dirac.ac.uk}).

\section*{Data availability}

The data underlying this article will be shared on reasonable request to the corresponding author.

%%%%%%%%%%%%%%%%%%%%%%%%%%%%%%%%%%%%%%%%%%%%%%%%%%



\bibliographystyle{mnras}
\bibliography{biblioteca} % if your bibtex file is called example.bib

\bsp	% typesetting comment
\label{lastpage}
\end{document}